\documentclass[%
 reprint,
 superscriptaddress,
 groupedaddress,
 unsortedaddress,
 runinaddress,
 frontmatterverbose, 
 showpacs,preprintnumbers,
 nofootinbib,
 nobibnotes,
 amsmath,amssymb,
 aps,
 pra,
 prb,
 rmp,
 prstab,
 prstper,
 floatfix,
 ]{revtex4-1}
\usepackage{amsmath}
\usepackage{amssymb}
\usepackage{units}
\usepackage{color}
\usepackage{graphicx}
\usepackage{dcolumn}
\usepackage{bm}
\usepackage{fixmath}
\usepackage{float}

\begin{document}


\title{Treatment of Herzberg-Teller and non-Condon effects in optical spectra with Hierarchical Equations of Motion}%

\author{Joachim Seibt}
\email{seibt@karlov.mff.cuni.cz}
%
%
%
\author{Tom\'{a}\v{s} Man\v{c}al}%
\affiliation{%
Faculty of Mathematics and Physics, Ke Karlovu 5, 121 16 Prague 2, Czech Republic
}%
%
%


\begin{abstract}
We derive a Hierarchical Equations of Motion (HEOM) description of nonadiabatic Herzberg-Teller type coupling effects and of non-Condon effects in a system of electronic transitions interacting with intra- and inter-molecular vibrational modes. We point out analogies between the auxiliary density operators (ADOs) of the hierarchy and the terms arising from explicit description of the vibrational modes in the Hamiltonian. In comparison with the standard formulation of HEOM, our equations contain additional connections between ADOs in the hierarchy scheme, which account for the dependence of the nonadiabatic coupling elements on the vibrational coordinates. We compare absorption spectra calculated with our HEOM methodology and with explicit treatment of vibrational DOF for a dimer system in the strong resonance coupling regime. Furthermore, we show that for sufficiently strong excitonic coupling, the corresponding effects in linear absorption spectra of vibronic dimers can be captured very well by the concept of effective Huang-Rhys factors.
\end{abstract} 

\maketitle


\section{Introduction}

In molecular aggregates, such as photosynthetic light harvesting complexes, nuclear degrees of freedom (DOF) play a substantial role in energy transfer processes subsequent to electronic excitation. Nuclear DOF, attributed mostly to the environment of the aggregate, determine the rates of dissipative processes, and allow thus the excitation to cross excitonic energy gaps 
\cite{AmVaGr00, May04,VaAbMa13, YaFl02_CP_355,JaChReEa08_JCP_101104,ChZhCh13_JCP_224112,SuFuIs16_JCP_204106,SeMa17_JCP_174109}. Recently,  
much attention has been devoted to the role of intramolecular vibrations in exciton dynamics. Intramolecular vibrational states were proposed to enhance excitation energy transfer under various conditions, most often through augmenting resonance conditions between electronic states and extending excited state delocalization \cite{ChKaPuMa12_JPCB_7449,ChChKa13_SR,KrKr12_JPCL_2828,OvPrCh16_PRA_020102,TiPeJo13_PNAS_1203,TiPeJo17_JCP_154308}. The effects of inter-pigment coupling and those of electron-phonon coupling with environmental nuclear modes (phonons) are to a large extent in competition. While the former leads to delocalization of electronic states, the later tends to localize electronic states through local energy gap fluctuations. 

Electronic DOF of molecular aggregates are well described by Frenkel exciton model \cite{SiCa94, AmVaGr00, May04}. In principle, all components of the Frenkel exciton Hamiltonian can be dependent on nuclear DOF, and such a nuclear dependence of exciton parameters may significantly change exciton dynamics \cite{LeTr17_JCP_075101}. The dependence of resonance coupling between transitions on different molecules on nuclear DOF, in particular on intramolecular vibrations, is known as Herzberg-Teller effect \cite{HeTe33_ZPC_410}. Nonadiabatic couplings of Herzberg-Teller type can be quantified for a specific molecular aggregate by electronic structure methods \cite{BaBlBa13_JCTC_4097,RoMiPe16_PCCP_8701}. In many cases, resonance coupling between molecular transitions can be described by dipole-dipole approximation. Most often, transition dipole moments are assumed independent of nuclear coordinates (Condon approximation). However, transition dipole moments dependent on nuclear coordinates translate this dependence on the resonant coupling, and non-Condon effects are therefore intimately connected to Herzberg-Teller effects, sometimes being both collected under one of the two terms.

While the involvement of intramolecular nuclear vibrational modes in spectra and energy transfer (e.g. enhancement of their associated beating amplitude in time-resolved spectra) requires some resonance between electronic and vibrational energy gaps (see e.g. \cite{ChKaPuMa12_JPCB_7449,ChChKa13_SR,TiPeJo13_PNAS_1203}), enhancement of Huang-Rhys factors of a mode involved in Herzberg-Teller coupling (further on in this work, we will call such modes Herzberg-Teller active) does not rely on any such resonance. The theories involving resonance have a problem with selectivity, because the resonance conditions for enhancement seem to be rather broad \cite{ChChKa13_SR,TiPeJo13_PNAS_1203}. Correspondingly, it is difficult to explain why only a small number of modes (out of the large number potentially present) seem to be enhanced in beating sensitive spectroscopy. It is possible that Herzberg-Teller coupling is more selective, because it requires specific molecular structure, rather than broad resonance between states.  

In the context of time-resolved experiments on photosynthetic complexes, Herzberg-Teller nonadiabatic couplings were proposed as a potential source of pronounced vibrational coherence beatings in two-dimensional electronic spectra of chlorosome of green sulfur bacteria \cite{DoMaVa14_JCP_115103}. In a dense, disordered electronic band of chlorosome, there is very little chance to enhance exclusively only the two observed vibrational frequencies purely via resonance effect. The Herzberg-Teller coupling provides a possible explanation for the beating amplitude which is larger than expected from the Huang-Rhys factors of the involved monomers \cite{DoMaVa14_JCP_115103}. Investigation of Herzberg-Teller coupling can thus potentially shed some more light on the character of coherent beating observed in time-resolved optical experiments on excitonic systems.

A number of theoretical methods has been used in the past for the description of nonadiabatic coupling effects in general and of their influence on transfer properties. These include semi-classical or mixed quantum classical methods, \cite{Thoss00JCP,KaSuSc18_JCP_2018}, wavefunction based \cite{AlHaEn17_JCP_064302} or density matrix based propagation \cite{KrGeDoJPB_12401914}, and Hierarchical Equations of Motion (HEOM) method \cite{DuTh16_JCPL_382}. Specific description of Herzberg-Teller effects and their spectroscopic signatures can be made in a vibronic basis \cite{GuZuBe09_JCP_154302,SiJe12_JCP_064111,MaZhLi_JCP_2014,LeTr17_JCP_075101}, i.e. with an explicit inclusion of the relevant (Herzberg-Teller active) vibrational modes into the Hamiltonian. However, as an explicit treatment of vibrational modes quickly increases system size, it becomes important to search for reduction techniques, i.e. for methods to include even the Herzberg-Teller active modes into the bath which is not explicitly propagated. To achieve this, influence of Herzberg-Teller effects on spectroscopic signals was studied by a line-shape-function based approach \cite{TaMu93_PRE_118}, which  was used for a theoretical investigation of signatures of non-Condon effects in two-dimensional spectra with an additional Fourier transformation with respect to the waiting time (so-called 3D-spectra) \cite{BiCaGe17_JCP_084311}. 
Recently, the so-called Dissipation Equations of Motion (DEOM) method \cite{ZhQiXu16_JCP_204109} was formulated in a way which allows to take non-Condon effects into account. The same work also studied the excited state dynamics of an excitonic dimer with electronic excitation of the monomer units beyond the Condon approximation. However, excitonic coupling independent of vibrational coordinates was assumed.
To continue the efforts towards formulating an efficient reduced density matrix description of Herzberg-Teller effects, we show in our article that both Herzberg-Teller and non-Condon effects in the dynamics of reduced density matrix can be described by HEOM, where the nuclear modes modulating resonance coupling are treated as a part of the bath. 
HEOM has recently become a very popular tool for investigation of energy transfer processes in photosynthetic aggregates. It is numerically exact, and it has been applied extensively in the recent years to mid-sized molecular aggregates, such as the Fenna-Mathews-Olson (FMO) complex \cite{Ishizaki05JPCJ,Tanimura06JPCJ,ShChNa09_JCP_164518,ShChNa09_JCP_084105,Chen09JCP,Chen10JCP,KrKr12_JPCL_2828,HeKrKr12_NJP_023018,StSc12_JCTC_2808,LiZhBa14_JCP_134106,OlKr14_JCP_164109,TaOu15_JCP_224112,WiDa15_JCTC_3411,ScIv15_PR_1,XuSo17_JCP_064102,DiPr17_JCP_064102}. 
In our treatment of Herzberg-Teller and non-Condon effects we point out analogies between HEOM and description in a vibronic basis, which rely on the possibility to interpret the Auxiliary Density Operators (ADOs) of the Kubo-Tanimura hierarchy \cite{Tanimura06JPCJ} as a representation of a (stochastic) vibrational coordinate. This analogy was reported previously in a different context \cite{ShChNa09_JCP_164518,LiZhBa14_JCP_134106}. 
We show how the respective HEOM description can be derived, starting from Feynman-Vernon functional in analogy to Ref.~\cite{Tanimura06JPCJ}. We compare the results of the HEOM calculations with those of density matrix propagation in a vibronic basis and investigate the influence of Herzberg-Teller and non-Condon effects on absorption spectra of a dimer model.
This article is organized as follows: In Section \ref{sec:theoretical_background} we introduce the model system. We specify its treatment by density matrix propagation in a vibronic basis and by HEOM in Sections \ref{sec:density_matrix_propagation} and \ref{sec:HEOM}, respectively, and we describe the calculation of absorption spectra. In Section \ref{sec:results} we compare the absorption spectra from calculations with both approaches and discuss the influence of Herzberg-Teller coupling and non-Condon contributions to the transition dipole moments.

\section{Theoretical background} \label{sec:theoretical_background}

We start with the general formulation of the aggregate Hamiltonian, 
where the $m$-th monomer unit is characterized by creation and annihilation operators in the electronic basis $\hat{B}^{\dagger}_m$ and $\hat{B}_m$, and
electronic excitation energy $\epsilon_m$. We consider one intramolecular vibrational mode per monomer with  momentum operator $\hat{p}_m$, vibrational frequency $\omega_m$, 
reorganization energy $\lambda_m$ and position operator $\hat{q}_m$. In the aggregate, the monomers with indices $m$ and $n$ are connected via resonance (or excitonic) coupling $J_{mn}$.
We also assume that electronic excitations involve interaction with a large thermodynamic bath which causes dephasing of the electronic transitions on the monomers, and we include description of this external bath in the HEOM
approach in the standard way \cite{LiZhBa14_JCP_134106,Chen10JCP}.
The bath modes $\{ \alpha \}$ are characterized by momentum operator $\hat{p}_{\alpha}$, position operator $\hat{x}_{\alpha}$ and frequency $w_{\alpha}$.
In our HEOM treatment, we intentionally want to treat also the intramolecular vibrational mode as a part of the bath.
Correspondingly, we split the aggregate Hamiltonian into a system component $\hat{H}_{S}$, bath components $\hat{H}_{B}$ and $\hat{H}'_B$ and system-bath coupling components $\hat{H}_{SB}$ and $\hat{H}'_{SB}$, where the Hamiltonian contributions without a prime are related to the single oscillator mode associated with intramolecular vibrations, whereas those with a prime are related to the thermal bath. The decomposition of the total Hamiltonian
\begin{equation} 
\begin{split}
&\hat{H}=\hat{H}_{S}+\hat{H}_{B}+\hat{H}_{SB}+\hat{H}'_B+\hat{H}'_{SB} \\
&=\sum_{m} \epsilon_m \hat{B}^{\dagger}_m \hat{B}_m +\sum_{m} \sum_{n \neq m} J_{mn} \hat{B}^{\dagger}_m \hat{B}_n \\
&+\sum_m \frac{1}{2} \left[ \hat{p}_m^2 + \omega_m^2 \left( \hat{q}_m - \frac{\sqrt{2 \lambda_m}}{\omega_m} \hat{B}^{\dagger}_m \hat{B}_m \right)^2 \right] \\
&+\sum_{m} \sum_{\alpha} \frac{1}{2} \left[ \hat{p}_{\alpha}^2 + w_{\alpha}^2 \left( \hat{x}_{\alpha} - \frac{\sqrt{2 \lambda_{\alpha}}}{w_{\alpha}} \hat{B}^{\dagger}_m \hat{B}_m \right)^2 \right]
\label{eq:Hamiltonian_general}
\end{split}
\end{equation}
allows us to identify the different contributions, where
\begin{equation} \label{eq:Hamiltonian_general_system}
\begin{split}
\hat{H}_{S}&=\sum_{m} \left( \epsilon_m+\lambda_m +\sum_{\alpha} \lambda_{\alpha} \right) \hat{B}^{\dagger}_m \hat{B}_m \\
&+\sum_{m} \sum_{n \neq m} J_{mn} \hat{B}^{\dagger}_m \hat{B}_n,
\end{split}
\end{equation}
\begin{equation} \label{eq:Hamiltonian_general_bath}
\hat{H}_{B}=\sum_m \frac{1}{2} \hat{p}_m^2 + \sum_m \frac{1}{2} \omega_m^2 \hat{q}_m^2
\end{equation}
and
\begin{equation} \label{eq:Hamiltonian_general_system-bath}
\hat{H}_{SB}=-\sum_m \sqrt{2 \lambda_m} \omega_m \hat{q}_m \hat{B}^{\dagger}_m \hat{B}_m
\end{equation}
characterize the combination of electronic system (including reorganization energies) and associated vibrational mode, whereas the contributions
\begin{equation} \label{eq:Hamiltonian_bath}
\hat{H}'_B=\sum_{\alpha} \frac{1}{2} \left( \hat{p}^2_{\alpha}+w^2_{\alpha} \hat{x}^2_{\alpha} \right)
\end{equation}
and
\begin{equation} \label{eq:Hamiltonian_system_bath_coupling}
\hat{H}'_{SB}=-\sum_{\alpha} \sqrt{2 \lambda_{\alpha}} w_{\alpha} \hat{B}^{\dagger}_m \hat{B}_m \hat{x}_{\alpha}
\end{equation}
characterize the thermal bath.
The treatment of the Hamiltonian, Eq.~(\ref{eq:Hamiltonian_general}), is well-known both in the context of HEOM, as well as that of the density matrix propagation. In this paper we study Herzberg-Teller coupling terms originating via Taylor expansion of the resonance coupling dependence on intramolecular coordinates.  We define the first and second order Herzberg-Teller coupling constants $J^{(1)}_{mn}$ and $J^{(2)}_{mn}$ and extend the Hamiltonian, Eq. (\ref{eq:Hamiltonian_general}), by two terms
\begin{equation} \label{eq:nonadiabatic_coupling_first_order}
\hat{H}^{(1)}=\sum_{m} \sum_{n \neq m} J^{(1)}_{mn} (\hat{q}_m + \hat{q}_n) \hat{B}^{\dagger}_m \hat{B}_n
\end{equation}
and
\begin{equation} \label{eq:nonadiabatic_coupling_second_order}
\hat{H}^{(2)}=\sum_{m} \sum_{n \neq m} \frac{1}{2} J^{(2)}_{mn} (\hat{q}_m + \hat{q}_n)^2 \hat{B}^{\dagger}_m \hat{B}_n.
\end{equation}
For a general formulation of the Herzberg-Teller coupling, one needs to define mode coupling constants, which in Eqs.~(\ref{eq:nonadiabatic_coupling_first_order}) and (\ref{eq:nonadiabatic_coupling_second_order}) are taken as independent of which monomer unit the respective vibrational modes are assigned to. We choose this simple dependence on the bath coordinates here in order to reduce the number of model parameters. However, as will be shown below, contributions of terms depending on the squares of the coordinates and on the product of two different coordinates can be easily identified in the final formulation of the HEOM equations, and thus the result can be easily applied to a more general coupling.
Treatment of these terms requires a closer consideration in the case of HEOM. Even though drawing analogies from the description in a vibronic basis already allows us to reveal the structure of the corresponding contributions in HEOM,
a detailed derivation is required for a proper formulation.
The same holds for the description of non-Condon effects, which can be taken into account by supplementing the transition dipole operator in Condon approximation,
\begin{equation} \label{eq:transition_dipole_moment_Condon_term}
\hat{\vec{\mu}}^{(0)}=\sum_{m} \vec{\mu}^{(0)}_{m} (\hat{B}^{\dagger}_m + \hat{B}_m),
\end{equation}
by the non-Condon term
\begin{equation} \label{eq:transition_dipole_moment_non_Condon_term}
\hat{\vec{\mu}}^{(1)}=\sum_{m} \vec{\mu}^{(1)}_{m} \hat{q}_m (\hat{B}^{\dagger}_m + \hat{B}_m).
\end{equation}

\subsection{Density matrix propagation} \label{sec:density_matrix_propagation}
In order to numerically verify our HEOM treatment of the interaction of electronic DOF with the single displaced harmonic oscillator per monomer, Eq. (\ref{eq:Hamiltonian_general_bath}), we solve the problem defined above by explicitly propagating the master equation for the reduced density matrix of the system defined by the Hamiltonian $\hat{H}_{S}+\hat{H}_{B}+\hat{H}_{SB}$. The thermal bath in which the system is embedded is described by Hamiltonian $\hat{H}'_B$, and the Hamiltonian $\hat{H}'_{SB}$ describes the weak system--bath interaction. In the limit in which the system--bath coupling is truly weak, the reduced density matrix propagation should lead to the same results as our HEOM treatment, in which, however, bath is defined by Hamiltonian operator $\hat{H}_{B}+\hat{H}'_B$ and system--bath interaction by operator $\hat{H}_{SB}+\hat{H}'_{SB}$.  
For the formulation of the problem we use basis representation analogous to Ref. \cite{BaBuVa14_JCP_95}.

For the $m$-th monomer unit the ground state Hamiltonian can be identified with the $m$-th component of the bath term from Eq.~(\ref{eq:Hamiltonian_general_bath}),
\begin{equation} \label{eq:Hamiltonian_monomer_ground_state}
\hat{h}_m = \frac{\hat{p}^2_m}{2} + \frac{\omega^2_m}{2} \hat{q}^2_m,
\end{equation}
which by representing position and momentum operator in terms of bosonic creation and annihilation operators $\hat{b}^{\dagger}_m$ and $\hat{b}_m$ as $\hat{p}_m = i \sqrt{\frac{\omega_m}{2}}(\hat{b}^{\dagger}_m-\hat{b}_m)$
and $\hat{q}_m = \sqrt{\frac{1}{2 \omega_m}}(\hat{b}^{\dagger}_m+\hat{b}_m)$ becomes equivalent to
\begin{equation} \label{eq:Hamiltonian_monomer_ground_state_creation_and_annihilation_operators}
\hat{h}_m = \omega_m \left( \hat{b}^{\dagger}_m \hat{b}_m + \frac{1}{2} \right).
\end{equation}
By introducing the Huang-Rhys factor $S_m$, which can be related to reorganization energy $\lambda_m$ and displacement $d_m$ 
via $\lambda_m=\omega_m S_m$ and $d_m=\sqrt{\frac{2 S_m}{\omega_m}}$, respectively,
the so-called shift operator $\hat{D}_m=\exp(-\sqrt{S_m}(\hat{b}^{\dagger}_m-\hat{b}_m))$ can be defined \cite{BaBuVa14_JCP_95}.
The respective transformation accounts for the shifted equilibrium position of the nuclei in the singly excited state, as described by the system-bath coupling term in Eq.~(\ref{eq:Hamiltonian_general_system-bath}),
and for the reorganization energy contribution from the system component in Eq.~(\ref{eq:Hamiltonian_general_system}).
Furthermore, the electronic excitation energy is included, so that the monomer Hamiltonian reads:
\begin{equation} \label{eq:Hamiltonian_monomer}
\hat{H}_m = \hat{h}_m + \left( \epsilon_m + \hat{D}_m \hat{h}_m \hat{D}^{\dagger}_m \right) \hat{B}^{\dagger}_m \hat{B}_m.
\end{equation}
The aggregate Hamiltonian is obtained by including Coulomb coupling $J_{mn}$ between the monomer units $m$ and $n$
\begin{equation} \label{eq:Hamiltonian_aggregate}
\hat{H}_{el-vib} = \sum_{m} \hat{H}_m +\sum_{m} \sum_{n \neq m} J_{mn} \hat{B}^{\dagger}_m \hat{B}_n.
\end{equation}

In a representation with assignment of vibrational states $|\boldsymbol{i}\rangle$ and $|\boldsymbol{j}\rangle$ in a selected electronic state to each monomer \cite{BuVaAb14_JCP}, the Hamiltonians of the subspaces of ground- and singly excited state can be expressed as
\begin{equation}
\begin{split}
&\hat{H}^{g g}_{el-vib, \boldsymbol{i},\boldsymbol{j}} = \sum_l \langle \boldsymbol{i} | \hat{h}_l | \boldsymbol{j} \rangle \\
&= \left[ \sum_{l} \omega_l \left( i_l +\frac{1}{2} \right) \right] \delta_{\boldsymbol{i} \boldsymbol{j}},
\end{split}
\end{equation}
and
\begin{equation}
\begin{split}
&\hat{H}^{e_m e_n}_{el-vib, \boldsymbol{i},\boldsymbol{j}} = \delta_{m n} \Bigg[ \epsilon_m \delta_{\boldsymbol{i} \boldsymbol{j}}
+\sum_{l \neq m} \langle \boldsymbol{i} | \hat{h}_l | \boldsymbol{j} \rangle \\
&+\langle \boldsymbol{i} | \hat{D}_m \hat{h}_m \hat{D}^{\dagger}_m | \boldsymbol{j} \rangle \Bigg]
+ (1-\delta_{m n}) J_{m n} \delta_{\boldsymbol{i} \boldsymbol{j}}
\end{split}
\end{equation}
with the definitions $\delta_{\boldsymbol{i} \boldsymbol{j}}=\prod_{k} \delta_{i_k j_k}$.
In the vibronic basis the position coordinate is represented as
\begin{equation} \label{eq:creation_annihilation_boson_basis_representation}
\begin{split}
&\langle \boldsymbol{i} | \hat{q}_m | \boldsymbol{j} \rangle=\sqrt{\frac{1}{2 \omega_m}} \langle \boldsymbol{i} | (\hat{b}_m+\hat{b}^{\dagger}_m) | \boldsymbol{j} \rangle \\
&=\sqrt{\frac{1}{2 \omega_m}} \left( \sqrt{i_m+1}\delta_{i_{m+1} j_{m}}+\sqrt{i_m}\delta_{i_{m-1} j_{m}} \right).
\end{split}
\end{equation}

As pointed out in earlier works \cite{GuZuBe09_JCP_154302,SiJe12_JCP_064111,MaZhLi_JCP_2014}, coordinate-dependent coupling in the singly excited state of a dimer can be included in the
chosen basis representation in a straightforward way. For rescaling of the respective coupling constants to energy units, we introduce
$\tilde{J}^{(1)}=J^{(1)} \sqrt{\frac{1}{2 \omega_0}}$ and $\tilde{J}^{(2)}=J^{(2)} \frac{1}{2 \omega_0}$ under the assumption $\omega_0=\omega_m=\omega_n$.
\begin{widetext}
Then in addition to the already defined components a first-order term
\begin{equation} \label{eq:J_first_order}
\begin{split}
&\hat{H}^{(1)}_{el-vib, \boldsymbol{i},\boldsymbol{j}}=\sum_{m} \sum_{n \neq m} 
\langle \boldsymbol{i} | J^{(1)}_{mn} (\hat{q}_m+\hat{q}_n) \hat{B}^{\dagger}_m \hat{B}_n | \boldsymbol{j} \rangle
=\sum_{m} \sum_{n \neq m} \tilde{J}^{(1)}_{mn} \hat{B}^{\dagger}_m \hat{B}_n \\
&\times \left( (\sqrt{i_m+1}\delta_{i_{m+1} j_{m}}+\sqrt{i_m}\delta_{i_{m-1} j_{m}}) \prod_{k; k \neq m} \delta_{i_k j_k}
+(\sqrt{i_n+1}\delta_{i_{n+1} j_{n}}+\sqrt{i_n}\delta_{i_{n-1} j_{n}}) \prod_{k; k \neq n} \delta_{i_k j_k} \right)
\end{split}
\end{equation} 
and a second-order term
\begin{equation} \label{eq:J_second_order}
\begin{split}
&\hat{H}^{(2)}_{el-vib, \boldsymbol{i},\boldsymbol{j}}=\sum_{m} \sum_{n \neq m} \frac{1}{2} \langle \boldsymbol{i} | J^{(2)}_{mn} 
(\hat{q}_m+\hat{q}_n)^2 \hat{B}^{\dagger}_m \hat{B}_n | \boldsymbol{j} \rangle
=\sum_{m} \sum_{n \neq m} \frac{1}{2} \tilde{J}^{(2)}_{mn}  \hat{B}^{\dagger}_m \hat{B}_n \\
& \times \Bigg((\sqrt{i_m+2} \sqrt{i_m+1}\delta_{i_{m+2} j_{m}}+i_m \delta_{i_m j_{m}} 
+(i_m+1) \delta_{i_m j_{m}}+\sqrt{i_m-1}\sqrt{i_m}\delta_{i_{m-2} j_{m}}) \prod_{k; k \neq m} \delta_{i_k j_k} \\
&+(\sqrt{i_n+2}\sqrt{i_n+1}\delta_{i_{n+2} j_{n}}+i_n\delta_{i_n j_{n}}
+(i_n+1)\delta_{i_n j_{n}}+\sqrt{i_n-1}\sqrt{i_n}\delta_{i_{n-2} j_{n}}) \prod_{k; k \neq n} \delta_{i_k j_k} \\
&+2 (\sqrt{i_m+1}\delta_{i_{m+1} j_{m}}+\sqrt{i_m}\delta_{i_{m-1} j_{m}}) \prod_{k; k \neq m} \delta_{i_k j_k}
\times (\sqrt{i_n+1}\delta_{i_{n+1} j_{n}}+\sqrt{i_n}\delta_{i_{n-1} j_{n}}) \prod_{k; k \neq n} \delta_{i_k j_k} \Bigg)
\end{split}
\end{equation}
enter in the Hamiltonian.
\end{widetext}
The transition dipole moments are defined as
\begin{equation} \label{eq:mue_zeroth_order}
\hat{\vec{\mu}}^{(0)}_{el-vib, \boldsymbol{i},\boldsymbol{j}}=\sum_m \vec{\mu}_m^{(0)} (\hat{B}^{\dagger}_m + \hat{B}_m) \delta_{\boldsymbol{i} \boldsymbol{j}},
\end{equation}
leading to electronic transitions facilitated by interaction of the electric field of incident light with the electronic transition dipole vector $\vec{\mu}_m^{(0)}$.
A possible dependence of the transition dipole moments on the position coordinate can be included in analogy to Eq.~(\ref{eq:J_first_order}).
We define the rescaled non-Condon contribution to the transition dipole moment as $\vec{\tilde{\mu}}^{(1)}_m=\vec{\mu}^{(1)}_m \sqrt{\frac{1}{2 \omega_m}}$, and write
\begin{equation}  \label{eq:mue_first_order}
\begin{split}
&\hat{\vec{\mu}}^{(1)}_{el-vib, \boldsymbol{i},\boldsymbol{j}}=\sum_m \langle \boldsymbol{i} | \vec{\mu}_m^{(1)} \hat{q}_m (\hat{B}^{\dagger}_m + \hat{B}_m) | \boldsymbol{j} \rangle \\
&=\sum_m \vec{\tilde{\mu}}_m^{(1)} (\hat{B}^{\dagger}_m + \hat{B}_m) \\
& \times (\sqrt{i_m+1}\delta_{i_{m+1} j_{m}}+\sqrt{i_m}\delta_{i_{m-1} j_{m}}) \prod_{k; k \neq m} \delta_{i_k j_k}. 
\end{split}
\end{equation}

For the description of the bath we choose a Debye-Drude spectral density with a prefactor $\eta_{DD}$ corresponding to twice the reorganization energy
and with a so-called cutoff frequency $\omega_c$ corresponding to a damping constant. The Debye-Drude spectral density is of the form
\begin{equation} \label{eq:Debye-Drude_spectral_density}
J_{DD}(\omega)=\eta_{DD} \omega \frac{\omega_c}{\omega^2+\omega_c^2}.
\end{equation}
It can be inserted in the general formula for calculation of the correlation function of the bath component of $\hat{H}'_{SB}$ via
\begin{equation} \label{eq:correlation_function}
\begin{split}
&C_{el}(t)=\frac{2}{\pi} \int^{\infty}_{0} d \omega J_{DD}(\omega) \\
&\times \left( \cos(\omega t) \coth \left( \frac{\omega}{2 k_B T} \right) -i \sin(\omega t) \right).
\end{split}
\end{equation}
%
In the Supplementary Material we describe the treatment of relaxation by the standard Redfield approach in the framework of a vibronic basis representation in the time domain. 

Note that the secular Redfield approach is appropriate for the calculation of linear absorption if coherence transfer effects are negligible. If non-secular effects are significant, the nonsecular version of Redfield relaxation at least captures correct tendencies \cite{OlKr14_JCP_164109}. In the calculation of nonlinear spectra, which involve excited state dynamics, the standard Redfield treatment as such is not appropriate for an accurate description \cite{Ta15_JCP_144110}.

\subsection{Hierarchical Equations of Motion} \label{sec:HEOM}

\subsubsection{Matsubara decomposition of the correlation function}

In the HEOM description all harmonic oscillator contributions of the Hamiltonian, also those which are attributed to intramolecular vibrations, are taken into account as bath components. 
Contributions from the respective bath components can be decomposed according to the Matsubara scheme, so that coefficients $c_k$ and time-dependent factors $\exp(-\gamma_k t)$ containing frequencies $\gamma_k$ enter in the correlation function
\begin{equation} \label{eq:Matsubara_decomposition_correlation_function}
C(t)=\sum_{k} c_k \exp(-\gamma_k t).
\end{equation}
This general expression is applicable even in the case of bath components with temperature-independent frequencies $\gamma_k$, such as undamped oscillators which we discuss in the Supplementary Material in more detail. To avoid confusion with the so-called ``Matsubara frequencies'' $\gamma_k=\frac{2 \pi k}{\beta}$ with $\beta=\frac{1}{k_B T}$, which depend on temperature by definition and enter, for example, in the Matsubara decomposition of a Debye-Drude spectral density, we will use the more general notation ``Matsubara decomposition coefficients'' and ``Matsubara decomposition frequencies'' in the following.
In the case of undamped oscillations the correlation function can be formulated as
\begin{equation} \label{eq:coefficients_undamped_oscillators}
\begin{split} 
C(t)&=\frac{S_{UO} \omega_{UO}^2}{2} \\
& \times \Bigg( \exp(-i \omega_{UO} t) \left[ \coth\left(\frac{\beta \omega_{UO}}{2}\right)+1 \right] \\
&+\exp(+i \omega_{UO} t) \left[ \coth\left(\frac{\beta \omega_{UO}}{2}\right)-1 \right] \Bigg).
\end{split}
\end{equation}
From this formulation the Matsubara decomposition frequencies and the corresponding coefficients can be immediately identified. We will specify them explicitly later. 
Note that the Matsubara decomposition terms of the undamped oscillator correlation function correspond to the two lowest Matsubara decomposition terms of an underdamped (Brownian) oscillator \cite{Mukamel95} 
in the case of zero damping.

\subsubsection{Herzberg-Teller coupling terms with HEOM}

In this section, the derivation of the HEOM with involvement of Herzberg-Teller coupling is sketched. We take only undamped oscillators representing intramolecular vibrational modes into account in this derivation, as only these modes contribute to Herzberg-Teller coupling according to our assumption and thus lead to non-standard terms in the HEOM scheme. However, contributions of an environment, as described by a Debye spectral density contribution, can be easily included by extending the dimension of the hierarchy and by assigning the additional index digits of the ADOs to the corresponding Matsubara decomposition terms. Then contributions of the environment enter in the HEOM description via involvement of ADOs with change of a Matsubara index from the respective index segment.
Under the assumption that the initial density matrix $\hat{\rho}(t_0)$ corresponds to a product of system component $\hat{\rho}_s(t_0)$ and bath component $\hat{\rho}_b(t_0)$,
the time evolution of the reduced density matrix (i.e.~the density matrix with traced-out bath component) can be expressed using the time evolution operator in Hilbert space, $\hat{U}(t,t_0)=\exp(-i \int_{t_0}^{t} d\tau \hat{H})$, or the Liouville space evolution superoperator as
\begin{equation} \label{eq:time_evolution_density_matrix}
\begin{split}
\hat{\rho}(t)&=Tr_B \{ \hat{U}^{\dagger}(t,t_0) \hat{\rho}(t_0) \hat{U}(t,t_0) \} \\
&=\hat{\cal U}(t,t_0) \hat{\rho}_s(t_0).
\end{split}
\end{equation}
For the Liouville space time evolution superoperator we can write
\begin{equation} \label{eq:time_evolution_density_matrix_with_phase_factors_and_FV_functional}
\begin{split}
&\hat{\cal U}(t,t_0)=\int_{\alpha(t_0)}^{\alpha(t)} {\cal D} \alpha \int_{\alpha'(t_0)}^{\alpha'(t)} {\cal D} \alpha' \\ 
&\exp(i S[\alpha]) {\cal F}[\alpha,\alpha'] \exp(-i S[\alpha'])
\end{split}
\end{equation}
with action $S[\alpha]=\int_{t_0}^{t} d\tau \left[ \frac{1}{2}\dot{\alpha}^2(\tau)-(H_{el}+U_{pot}(\alpha(\tau))) \right]$ and the Feynman-Vernon functional \cite{Tanimura06JPCJ}
\begin{equation} \label{eq:Feynman_Vernon_functional_correlation_function_real_imag_general_expression}
\begin{split}
&{\cal F}(\alpha,\alpha';t)=\exp \left\{ -\int_{0}^t d\tau \hat{V}^{\times}(\alpha,\alpha';t) \right. \\
& \left. \times \left[ \frac{\partial}{\partial \tau} \int_{0}^{\tau} d\tau' i \bar{L}_1(\tau-\tau') \hat{V}^{\circ}(\alpha,\alpha';t) \right. \right. \\
& \left. \left. +\int_{0}^{\tau} d\tau' L_2(\tau-\tau') \hat{V}^{\times}(\alpha,\alpha';t) \right] \right\}.
\end{split}
\end{equation}
Here,  $\hat{V}(\alpha)$ represents a system part of the system-bath interaction operator, and  the definitions $\hat{V}^{\times}(\alpha,\alpha';t)=\hat{V}(\alpha)-\hat{V}(\alpha')$ and $\hat{V}^{\circ}(\alpha,\alpha';t)=\hat{V}(\alpha)+\hat{V}(\alpha')$, 
$\bar{L}_1(t)=\int_{0}^{\infty} d \omega \frac{J(\omega)}{\omega} \cos(\omega t)$ and
$L_2(t)=\int_{0}^{\infty} d \omega J(\omega) \coth \left( \frac{\omega}{2 k_B T} \right) \cos(\omega t)$ enter.

Based on the Feynman-Vernon functional, we derive the contributions to the HEOM scheme for an undamped oscillator with spectral density $J(\omega)=\frac{1}{2} S_{UO} \omega_{UO} \omega (\delta(\omega-\omega_{UO})+\delta(\omega+\omega_{UO}))$ in the Supplementary Material.
Note that in the case of an undamped oscillator only two terms appear in the Matsubara decomposition, and that the respective Matsubara decomposition frequencies are temperature-independent.
The derivation allows us to identify the Matsubara decomposition coefficients assigned to the Matsubara decomposition frequencies $\gamma_{1}=i \omega_{UO}$ and $\gamma_{2}=-i \omega_{UO}$ for the description of an undamped oscillator with HEOM as
\begin{eqnarray} \label{eq:Matsubara_coefficients_UO}
c_{1}&=&\frac{1}{2} S_{UO} \omega_{UO}^2 \left( \coth\left( \frac{\beta \omega_{UO}}{2} \right) +1 \right), \\
c_{2}&=&\frac{1}{2} S_{UO} \omega_{UO}^2 \left( \coth\left( \frac{\beta \omega_{UO}}{2} \right) -1 \right)
\end{eqnarray}
with $\tilde{c}_{1}=c_{2}$ and $\tilde{c}_{2}=c_{1}$ in agreement with Ref.~\cite{Ta12_JCP_22A550}.

In the case of a dimer, $\hat{V}$ can be separated into components $\hat{V}_l=\hat{B}_l^{\dagger} \hat{B}_l$. With this definition and by applying the Matsubara decomposition of the bath correlation function, the Feynman-Vernon functional can be reformulated as
\begin{equation}
\label{eq:FV-functional_reformulated}
\begin{split}
&{\cal F}(\alpha,\alpha';t)=\exp \Bigg\{ -\int_{0}^t d\tau \sum_l \sum_k \hat{\Phi}_l(\alpha,\alpha') \\
&\times \Bigg( \Bigg[ \int_{0}^{\tau} d\tau' \exp(-\gamma_k (\tau-\tau')) \hat{\Theta}_{lk}(\alpha,\alpha',t) \Bigg] \\
& +\hat{G}_{lk}(\alpha,\alpha',t) \Bigg)  \Bigg\}
\end{split}
\end{equation}
with
\begin{eqnarray} 
\label{eq:Phi}
\hat{\Phi}_l(\alpha,\alpha')&=&i \left[ \hat{V}_l(\alpha)-\hat{V}_l(\alpha') \right], \\
\label{eq:Theta}
\hat{\Theta}_{l k}(\alpha,\alpha',t)&=&-i \left[ c_k \hat{V}_l(\alpha) -\tilde{c}_k \hat{V}_l(\alpha') \right], \\
\label{eq:G_term}
\hat{G}_l(\alpha,\alpha',t)&=&S_{UO} \omega_{UO} \exp(-\gamma_k t) \nonumber \\
&& \times \left[ \hat{V}_l(\alpha) +\hat{V}_l(\alpha') \right].
\end{eqnarray} 
Different from the treatment of a Debye-Drude spectral density, $\hat{\Theta}_{l k}(\alpha,\alpha',t)$ does not exhibit an explicit time-dependence because of the purely real Matsubara decomposition coefficients of the undamped oscillator correlation function.
By repeated application of the recursion scheme which is explained in Supplementary Information, in combination with the definition from Eq.~(\ref{eq:time_evolution_density_matrix_with_phase_factors_and_FV_functional}), one obtains auxiliary density operators (ADOs) $\hat{\rho}_{\boldsymbol{n}}(t)=\int_{\alpha(t_0)}^{\alpha(t)} {\cal D} \alpha \int_{\alpha'(t_0)}^{\alpha'(t)} {\cal D} \alpha' \hat{\rho}_{\boldsymbol{n}}(\alpha,\alpha',t)$ with $\boldsymbol{n}=\{ \{ n_{11}, n_{12} \}, \{ n_{21}, n_{22} \} \}$ and
\begin{equation}
\begin{split}
&\hat{\rho}_{\boldsymbol{n}}(\alpha,\alpha',t)=\exp(i S[\alpha]) {\cal F}(\alpha,\alpha';t) \exp(-i S[\alpha']) \\
&\prod_l \prod_k \Bigg[ \int_{0}^{t} d\tau' \exp(-\gamma_k (t-\tau')) \hat{\Theta}_{l k}(\alpha,\alpha',t) \\ 
&+\hat{G}_{l k}(\alpha,\alpha',t) \Bigg]^{n_{lk}}.
\end{split}
\end{equation}
The equation of motion for this ADO is obtained as
\begin{equation} \label{eq:evaluation_HEOM_scheme_dimer_first_step}
\begin{split}
&\frac{\partial}{\partial t} \hat{\rho}_{\boldsymbol{n}}(\alpha,\alpha',t)=\Bigg( i (S[\alpha]-S[\alpha']) \\
& -i \sum_{l} \sum_{k} n_{lk} \gamma_k \Bigg) \hat{\rho}_{\boldsymbol{n}}(\alpha,\alpha',t) \\
&-\sum_l \sum_k \hat{\Phi}_{l}(\alpha,\alpha') \hat{\rho}_{\boldsymbol{n}^{+}_{lk}}(\alpha,\alpha',t) \\
&+ \sum_l \sum_k n_{lk} \hat{\Theta}_{lk}(\alpha,\alpha',t) \hat{\rho}_{\boldsymbol{n}^{-}_{lk}}(\alpha,\alpha',t),
\end{split}
\end{equation}
and it can be reformulated after the path integrations as
\begin{equation} \label{eq:evaluation_HEOM_scheme_dimer_second_step}
\begin{split}
&\frac{\partial}{\partial t} \hat{\rho}_{\boldsymbol{n}}=-\left( i \hat{\cal L} + \sum_{l} \sum_{k} n_{lk} \gamma_k \right) \hat{\rho}_{\boldsymbol{n}} \\
&-i \sum_l \sum_k \left[ \hat{B}_l^{\dagger} \hat{B}_l, \hat{\rho}_{\boldsymbol{n}^{+}_{lk}} \right] \\
&-i \sum_l \sum_k  n_{lk} \left(c_k \hat{B}_l^{\dagger} \hat{B}_l \hat{\rho}_{\boldsymbol{n}^{-}_{lk}}
-\hat{\rho}_{\boldsymbol{n}^{-}_{lk}} \tilde{c}_k \hat{B}_l^{\dagger} \hat{B}_l \right).
\end{split}
\end{equation}
After the rescaling 
$\hat{\tilde{\rho}}_{\boldsymbol{n}}=\prod_l \prod_k (|c_k|^{n_{lk}} n_{lk}!)^{-\frac{1}{2}} \hat{\rho}_{\boldsymbol{n}}$ (see Ref. \cite{ShChNa09_JCP_164518}) the resulting HEOM equation reads as
\begin{equation} \label{eq:evaluation_HEOM_scheme_dimer_rescaled}
\begin{split}
&\frac{\partial}{\partial t} \hat{\tilde{\rho}}_{\boldsymbol{n}}=-\left( i \hat{\cal L} + \sum_{l} \sum_{k} n_{lk} \gamma_k \right) \hat{\tilde{\rho}}_{\boldsymbol{n}} \\
&-i \sum_l \sum_k \sqrt{(n_{lk}+1) c_k} \left[ \hat{B}_l^{\dagger} \hat{B}_l, \hat{\tilde{\rho}}_{\boldsymbol{n}^{+}_{lk}} \right] \\
&-i \sum_l \sum_k \sqrt{\frac{n_{lk}}{c_k}} \left( c_k \hat{B}_l^{\dagger} \hat{B}_l \hat{\tilde{\rho}}_{\boldsymbol{n}^{-}_{lk}}
-\hat{\tilde{\rho}}_{\boldsymbol{n}^{-}_{lk}} \tilde{c}_k \hat{B}_l^{\dagger} \hat{B}_l \right).
\end{split}
\end{equation}
From this formulation the analogy between prefactors with square roots of Matsubara decomposition indices and prefactors with square roots of vibrational quantum numbers in the vibronic basis representation becomes recognizable. Moreover, ADOs with changing index digits can be interpreted in terms of influence of bosonic creation/annihilation operators in a representation of the position coordinate. On the basis of this finding, we will point out the possibility, but also the limitations, of drawing analogies between density matrix propagation in a vibronic basis and HEOM for the formulation of Herzberg-Teller couplings and non-Condon effects.
Note that in addition to the undamped oscillator a Debye-Drude spectral density can be taken into account in the HEOM scheme by extending the subscript set of Matsubara decomposition indices \cite{LiZhBa14_JCP_134106}. If ADOs with decreased Matsubara decomposition indices assigned to a Debye-Drude spectral density component become involved in the HEOM equations, instead of $\tilde{c}_k$ the complex conjugate coefficient $c^{*}_k$ appears in the respective terms \cite{Chen10JCP}.
  
The inclusion of nonadiabatic coupling terms (with respect to the undamped oscillator spectral density component) is not straightforward, unless they correspond to harmonic oscillators with shifted equilibrium position, which is not the case under the assumption that the nonadiabatic coupling is of Herzberg-Teller type. While off-diagonal system-bath coupling terms of shifted-harmonic-oscillator type could be treated by extending the hierarchy structure by Matsubara decomposition indices related to the off-diagonal coupling coordinate, in the present case a description relying on the existing hierarchy structure is possible. This approach leads to additional terms in the HEOM to account for the interplay between adjacent ADOs under the influence of Herzberg-Teller coupling. From the computational point of view, the fact that the existing hierarchy does not have to be extended is advantageous, as the number of included Matsubara decomposition terms appears as a factorial in the scaling of the required number of ADOs \cite{ShChNa09_JCP_084105}. However, the efficiency of HEOM calculations critically depends on the truncation scheme, i.e.~on the selection of ADOs with sufficiently large contribution to justify their involvement in the solution of the HEOM equations. The rescaling introduced above does not only allow an intuitive comparison with description in a vibronic basis, but also leads to improved numerical convergence. If the required order of hierarchy layers for a sufficiently accurate calculation would be independent of included Herzberg-Teller coupling, the numerical effort, as compared to density matrix propagation in a vibronic basis, would be the same as without such effects. Despite the superexponential scaling of the number of ADOs in the hierarchy structure with increasing number of involved monomer units and increasing number of hierarchy layers, the HEOM approach with appropriate hierarchy truncation is favorable over description in a vibronic basis even for large aggregates. However, an {\it a priori} statement about the influence of Herzberg-Teller effects on the convergence properties is difficult. 
In the calculation of dimer absorption spectra, where the effect of Herzberg-Teller coupling consists in modified Huang-Rhys factors according to our findings in Sec.~\ref{sec:results}, truncation of the hierarchy scheme seems to be possible at similar depth independent of whether Herzberg-Teller coupling is taken into account or the Huang-Rhys factors are modified accordingly instead.

The first-order Herzberg-Teller coupling Hamiltonian from Eq.~(\ref{eq:nonadiabatic_coupling_first_order}) can be taken into account in the Feynman-Vernon functional in terms of its time-dependent correlation with components of the system-bath coupling Hamiltonian,
thereby identifying the components of the system part associated with coupling between electronic states $l$ and $m$ as $\hat{V}_{J^{(1)},lm}=\hat{B}_l^{\dagger} \hat{B}_m$ and taking the trace only with respect to the bath.
To include these correlation functions of the form
\begin{eqnarray} \label{eq:correlation_function_H_SB_H_J1_first_equation}
&&\langle \hat{H}_{SB,l}(t) \hat{H}_{lm}^{(1)}(0) \rangle_{bath}=-\sqrt{S_{UO}} \omega_{UO} \tilde{J}_{lm}^{(1)}, \nonumber \\
&&\times \langle \hat{b}_l^{\dagger}(t)+\hat{b}_l(t),\hat{b}_l^{\dagger}+\hat{b}_l \rangle \hat{B}_l^{\dagger} \hat{B}_m, \\
\label{eq:correlation_function_H_SB_H_J1_second_equation}
&&\langle \hat{H}_{ml}^{(1)}(t) \hat{H}_{SB,l}(0) \rangle_{bath}=-\sqrt{S_{UO}} \omega_{UO} \tilde{J}_{ml}^{(1)} \nonumber \\
&&\times \langle \hat{b}_l^{\dagger}(t)+\hat{b}_l(t),\hat{b}_l^{\dagger}+\hat{b}_l \rangle \hat{B}_m^{\dagger} \hat{B}_l
\end{eqnarray}
in addition to the correlation function of system-bath coupling
\begin{equation} \label{eq:correlation_function_H_SB_H_SB}
\begin{split}
&\langle \hat{H}_{SB,l}(t) \hat{H}_{SB,l}(0) \rangle_{bath}= S_{UO} \omega_{UO}^2 \\
&\times \langle \hat{b}_l^{\dagger}(t)+\hat{b}_l(t),\hat{b}_l^{\dagger}+\hat{b}_l \rangle \hat{B}_l^{\dagger} \hat{B}_l \\
&=\left( \sum_k c_{k} \exp(-\gamma_k t) \right) \hat{B}_l^{\dagger} \hat{B}_l,
\end{split}
\end{equation}
an appropriate scaling factor is required to relate Eqs.~(\ref{eq:correlation_function_H_SB_H_J1_first_equation}) and (\ref{eq:correlation_function_H_SB_H_J1_second_equation}) to the given Matsubara decomposition of Eq.~(\ref{eq:correlation_function_H_SB_H_SB}).
Note that Eq.~(\ref{eq:correlation_function_H_SB_H_SB}) was formulated under the assumption that both monomer units are characterized by the same correlation function (otherwise separate Matsubara decompositions would lead to coefficients and frequencies with two indices).
The required scaling factor can be identified as
$-\frac{\tilde{J}_{lm}^{(1)}}{\sqrt{S_{UO}} \omega_{UO}}=-\frac{\tilde{J}_{lm}^{(1)}}{\sqrt{c_k}} \frac{\sqrt{c_k}}{\sqrt{S_{UO}} \omega_{UO}}=-\frac{\bar{J}_{lm,k}^{(1)}}{\sqrt{c_k}}$,
where the factor $\frac{\sqrt{c_k}}{\sqrt{S_{UO}} \omega_{UO}}$ corresponds to $\sqrt{\frac{\coth \left( \frac{\beta \omega_{UO}}{2} \right)+(-1)^k}{2}}$ and thus leads to a temperature-dependence of $\bar{J}_{lm}^{(1)}$.
Note that the concept of a temperature-dependent or ``dressed'' coupling -- however coupling of zeroth order, i.e.~coordinate-independent excitonic coupling -- also plays a role in the dissipative dynamics of open quantum systems with description in the polaron basis \cite{SiHa84_JCP_2615,ChZhCh13_JCP_224112}.
To include the first-order Herzberg-Teller term in the HEOM scheme, one can redefine Eqs.~(\ref{eq:Phi}), (\ref{eq:Theta}) and (\ref{eq:G_term}) by replacing $\hat{V}_l$ with $\hat{V}_l-\sum_m \frac{\bar{J}_{lm,k}^{(1)}}{\sqrt{c_k}} \hat{V}_{J^{(1)},lm}-\sum_m \frac{\bar{J}_{ml,k}^{(1)}}{\sqrt{c_k}} \hat{V}_{J^{(1)},ml}$, where the second and the third term stem from correlation functions formulated in Eqs.~(\ref{eq:correlation_function_H_SB_H_J1_first_equation}) and (\ref{eq:correlation_function_H_SB_H_J1_second_equation}), respectively.
\begin{widetext}
Derivation in analogy to Eqs.~(\ref{eq:evaluation_HEOM_scheme_dimer_first_step}) and (\ref{eq:evaluation_HEOM_scheme_dimer_second_step}) 
leads to the additional terms 
\begin{equation} \label{eq:evaluation_HEOM_scheme_dimer_second_step_J1}
\begin{split}
\left( \frac{\partial}{\partial t} \hat{\rho}_{\boldsymbol{n}} \right)_{\bar{J}^{(1)}}= \sum_l \sum_{m \neq l} \sum_k
i \frac{\bar{J}_{lm,k}^{(1)}}{\sqrt{c_k}}
&\left(
\left[ \hat{V}_{J^{(1)},lm}, \hat{\rho}_{\boldsymbol{n}^{+}_{lk}} \right]
+n_{lk} \left( c_k \hat{V}_{J^{(1)},lm} \hat{\rho}_{\boldsymbol{n}^{-}_{lk}}-\hat{\rho}_{\boldsymbol{n}^{-}_{lk}} \tilde{c}_k \hat{V}_{J^{(1)},lm} \right) \right. \\
&\left. +\left[ \hat{V}_{J^{(1)},lm}, \hat{\rho}_{\boldsymbol{n}^{+}_{mk}} \right]
+n_{mk} \left( c_k \hat{V}_{J^{(1)},lm} \hat{\rho}_{\boldsymbol{n}^{-}_{mk}}-\hat{\rho}_{\boldsymbol{n}^{-}_{mk}} \tilde{c}_k \hat{V}_{J^{(1)},lm} \right) \right).
\end{split}
\end{equation}
Application of the rescaling of the ADOs, thereby introducing redefined commutators
\begin{equation} \label{eq:redefinition_commutators}
\left[ \bullet,\hat{\tilde{\rho}}_{\boldsymbol{n}^{-}_{jk}} \right]_c=\bullet \hat{\tilde{\rho}}_{\boldsymbol{n}^{-}_{jk}} -\frac{\tilde{c}_k}{c_k} \hat{\tilde{\rho}}_{\boldsymbol{n}^{-}_{jk}} \bullet
\end{equation}
with appearance of a factor $\frac{\tilde{c}_k}{c_k}=\exp((-1)^{k} \beta \omega_{UO})$ in the subtracted term, however only in the case of involvement of an ADOs with a decreased Matsubara decomposition index, leads to
\begin{equation} \label{eq:evaluation_HEOM_scheme_dimer_second_step_J1_rescaling}
\begin{split}
\left( \frac{\partial}{\partial t} \hat{\tilde{\rho}}_{\boldsymbol{n}} \right)_{\bar{J}^{(1)}}= \sum_l \sum_{m \neq l} \sum_k 
i \bar{J}_{lm,k}^{(1)} 
&\Bigg(
\sqrt{n_{lk}+1} \left[ \hat{B}_l^{\dagger} \hat{B}_m, \hat{\tilde{\rho}}_{\boldsymbol{n}^{+}_{lk}} \right]
+\sqrt{n_{lk}} \left[ \hat{B}_l^{\dagger} \hat{B}_m, \hat{\tilde{\rho}}_{\boldsymbol{n}^{-}_{lk}} \right]_c \\
&+\sqrt{n_{mk}+1} \left[ \hat{B}_l^{\dagger} \hat{B}_m, \hat{\tilde{\rho}}_{\boldsymbol{n}^{+}_{mk}} \right]
+\sqrt{n_{mk}} \left[ \hat{B}_l^{\dagger} \hat{B}_m, \hat{\tilde{\rho}}_{\boldsymbol{n}^{-}_{mk}} \right]_c \Bigg).
\end{split}
\end{equation}
\end{widetext}
With increasing temperature the contributions which stem from different Matsubara decomposition terms, specified by the index $k$, become increasingly similar. For temperature approaching zero one finds $\bar{J}_{lm,1}^{(1)}=\unit[0]{}$ and $\bar{J}_{lm,2}^{(1)}=\tilde{J}_{lm}^{(1)}$, while the ratio $\frac{\tilde{c}_1}{c_1}$ diverges (even in combination with $\bar{J}_{lm,1}^{(1)}$) and $\frac{\tilde{c}_2}{c_2}$ approaches zero.
Note that by drawing analogies from the description in a vibronic basis (see Eq.~(\ref{eq:J_first_order})), one would expect a negative sign of the first-order Herzberg-Teller coupling terms, as excitonic coupling terms from the Liouville operator also enter with a negative sign.
The inverted sign appears because Eq.~(\ref{eq:evaluation_HEOM_scheme_dimer_second_step_J1_rescaling}) results from the combination of the opposite-signed first-order Herzberg-Teller coupling and system-bath coupling contributions within the Feynman-Vernon functional. Furthermore, drawing analogies from the description in a vibronic basis does not allow us to explain that instead of the Herzberg-Teller coupling constants $\tilde{J}_{lm}^{(1)}$ their thermally averaged equivalents $\bar{J}_{lm,k}^{(1)}$ enter.
It also becomes recognizable only from the analytical derivation that for terms including ADOs with decreasing index digits the redefined commutators from Eq.~(\ref{eq:redefinition_commutators}) with involvement of quotients of Matsubara decomposition coefficients in the subtracted term appear. However, the description of first-order Herzberg-Teller coupling in vibronic basis and HEOM exhibits analogous structure, where its representation in the vibronic basis can be related to the way how adjacent ADOs in the corresponding term from the HEOM approach are connected to each other. 

While the treatment of transition dipole contributions in Condon approximation is straightforward, 
a dependence of the transition dipole operator on vibrational coordinates in the case of non-Condon transition dipole contributions poses the task to formulate an equivalent expression in HEOM space. For the further derivation, the non-Condon contributions to the transition dipole moment are written as $\hat{\vec{\mu}}^{(1)}=\sum_l \vec{\xi}_{\mu^{(1)},l} \mu^{(1)}_l \hat{q}_l \hat{V}_{\mu^{(1)},l}$. We need to evaluate the expectation value of the non-Condon part of the transition dipole moment, i.e. $Tr_B \left\{ \mathbold{\hat{\vec{\mu}}}^{(1)} \hat{\rho}(t) \right\}$, where $\hat{\rho}(t)$ is the total density matrix (or any component of the perturbation series of the total density matrix to it) and the operator $\hat{V}_{\mu^{(0)},l}=\hat{V}_{\mu^{(1)},l}=\hat{B}_l^{\dagger}+\hat{B}_l$ facilitating the electronic excitation.
The expectation value can be written in terms of a generating function with an auxiliary time-like parameter $a$. We define an operator
\begin{equation}
\begin{split}
U_{\mu}(a,0) = \exp \Bigg\{ &-i \int_{0}^{a} d\tau H \\ 
&+ i \int_{0}^{a} d\tau \vec{\mu}^{(1)}E_{0} \Bigg\},
\end{split}
\end{equation}
which formally corresponds to the evolution operator with the total Hamiltonian $H$ and a light--matter interaction Hamiltonian with some arbitrary constant electric field $E_0$.
Now we can write for the non-Condon term
\begin{equation}
\begin{split}
&Tr_B \left\{ \mathbold{\hat{\vec{\mu}}}^{(1)} \hat{\rho}(t) \right\}=\sum_l i \frac{\mu^{(1)}_l}{E_0} \frac{\partial}{\partial \mu^{(1)}_l} \\
&\Bigg[  \frac{\partial}{\partial a} Tr_B \{ \hat{U}_{\mu}^{\dagger}(a,0) \hat{\rho}(t) \hat{U}_{\mu}(a,0) \} \Bigg|_{a=0}\Bigg]
\Bigg|_{\mu^{(1)}_l=0}. 
\end{split}
\end{equation}
The right-hand-side of the expression can be evaluated using ADOs from the hierarchy. In particular, we obtain
\begin{equation}
\label{eq:mu_in_hierarchy}
Tr_B \left\{ \mathbold{\hat{\vec{\mu}}}^{(1)} \hat{\rho}(t) \right\}= 
-\sum_l \sum_k \vec{\bar{\mu}}_{l,k}^{(1)} \left[ \hat{V}_{\mu^{(1)},l}, \hat{\rho}_{\boldsymbol{0}^{+}_{lk}} \right],
\end{equation}
where the index $\boldsymbol{0}=\{\{0,0\},\{0,0\}\}$ denotes the lowest term of the hierarchy, i.e. the reduced density matrix. In line with our previous definitions $\boldsymbol{0}^{+}_{lk}$ denotes hierarchy elements with one index equal to one. 
To evaluate expectation values of operators defined solely on the Hilbert space of the system, only the reduced density matrix is needed. For the calculation of expectation values of the operators with dependence on bath coordinates, such as $\hat{\vec{\mu}}^{(1)}$, HEOM provides elements of the hierarchy, in which the corresponding higher order information about the bath is kept. 

In order to express the action of the coordinate-dependent operators on the RDM or in fact any member of the hierarchy, we can introduce HEOM space operators such that their action on the member of the hierarchy is expressed in terms of a linear combination of (in general) all members of the hierarchy. We can define e.g.
\begin{equation}
\begin{split}
\label{eq:HEOM_space_1}
\mathbold{\vec{\mu}}^{(1)}_{H} \hat{\rho}_{\boldsymbol{0}} &= Tr_B \left\{ \mathbold{\hat{\vec{\mu}}}^{(1)} \hat{\rho}(t) \right\} \\ 
&= -\sum_l \sum_k \vec{\bar{\mu}}_{l,k}^{(1)} \left[ \hat{V}_{\mu^{(1)},l}, \hat{\rho}_{\boldsymbol{0}^{+}_{lk}} \right].
\end{split}
\end{equation}
For a calculation of the absorption spectrum, we need to evaluate action of $\tilde{\vec{\mu}}^{(1)}$ on the RDM, at time $t_0$, and at later time $t$ in which polarization is generated. However, for evaluation of higher-order signals and general spectroscopic signals, it is possible to establish HEOM space operators acting on arbitrary members of the hierarchy.

In the derivation how the influence of transitions is expressed in terms of hierarchical equations involving the ADOs equivalent to $\hat{\rho}(t)$, the Condon transition dipole contributions enter in the phase factors containing the action, whereas the non-Condon contributions enter in the Feynman-Vernon functional. 
Non-Condon contributions lead to redefinition of Eqs.~(\ref{eq:Phi}), (\ref{eq:Theta}) and (\ref{eq:G_term}) with replacement of $\hat{V}_l$ by $\hat{V}_l -\frac{\vec{\bar{\mu}}_{l,k}^{(1)}}{\sqrt{c_k}} \hat{V}_{\mu^{(1)},l}$ with the temperature-dependent non-Condon transition dipole contribution $\vec{\bar{\mu}}_{l,k}^{(1)}=\vec{\mu}_{l,k}^{(1)} \sqrt{\frac{\coth \left( \frac{\beta \omega_{UO}}{2} \right)+(-1)^k}{2}}$.
Accordingly, one obtains
\begin{equation} \label{eq:non-Condon_transition_dipole_moment}
\begin{split}
&\mathbold{\vec{\mu}}^{(1)}_{H} \hat{\rho}_{\boldsymbol{n}}= 
-\sum_l \sum_k \vec{\bar{\mu}}_{l,k}^{(1)} \Bigg( \left[ \hat{V}_{\mu^{(1)},l}, \hat{\rho}_{\boldsymbol{n}^{+}_{lk}} \right] \\
& +n_{lk} \left( \hat{V}_{\mu^{(1)},l} \hat{\rho}_{\boldsymbol{n}^{-}_{lk}}-\hat{\rho}_{\boldsymbol{n}^{-}_{lk}} \frac{\tilde{c}_k}{c_k} \hat{V}_{\mu^{(1)},l} \right) \Bigg).
\end{split}
\end{equation}
After switching to Hilbert space and rescaling the ADOs, a separation of the non-Condon contributions with transition dipole operator appearing on left- and right hand side of the density matrix leads to 
\begin{eqnarray} 
\label{eq:influence_non_Condon_term_lhs}
\hat{\mu}^{(1)}_{H} \hat{\tilde{\rho}}_{\boldsymbol{n}} &=&\sum_l \sum_k \left(
-\sqrt{n_{lk}+1} \vec{\bar{\mu}}_{l,k}^{(1)} (\hat{B}_l^{\dagger}+\hat{B}_l) \hat{\tilde{\rho}}_{\boldsymbol{n}^{+}_{lk}} \right. \nonumber \\
&&\left. -\sqrt{n_{lk}} \vec{\bar{\mu}}_{l,k}^{(1)} (\hat{B}_l^{\dagger}+\hat{B}_l) \hat{\tilde{\rho}}_{\boldsymbol{n}^{-}_{lk}} \right), \\
\label{eq:influence_non_Condon_term_rhs}
\hat{\tilde{\rho}}_{\boldsymbol{n}} \tilde{\vec{\mu}}^{(1)}_{H}&=&\sum_l \sum_k \left(
-\sqrt{n_{lk}+1} \hat{\tilde{\rho}}_{\boldsymbol{n}^{+}_{lk}} \vec{\bar{\mu}}_{l,k}^{(1)} (\hat{B}_l^{\dagger}+\hat{B}_l) \right. \nonumber \\
&&\left. -\sqrt{n_{lk}} \frac{\tilde{c}_k}{c_k} \hat{\tilde{\rho}}_{\boldsymbol{n}^{-}_{lk}} \vec{\bar{\mu}}_{l,k}^{(1)} (\hat{B}_l^{\dagger}+\hat{B}_l) \right).
\end{eqnarray}

A closer consideration of Eqs.~(\ref{eq:influence_non_Condon_term_lhs}) and (\ref{eq:influence_non_Condon_term_rhs}) leads to the following findings: As in the first-order Herzberg-Teller coupling, also in the non-Condon transition dipole moment the appearing signs are opposite to the ones which would be expected by drawing analogies from the corresponding formulation in the vibronic basis (see Eq.~(\ref{eq:mue_first_order})). Again, this finding can be explained by the opposite sign of the additional contribution (here: the interaction of the non-Condon transition dipole moment with the electric field) and the system-bath coupling. Furthermore, the appearance of the thermally averaged transition dipole moments $\vec{\bar{\mu}}_{l,k}^{(1)}$ instead of $\vec{\tilde{\mu}}_{l}^{(1)}$ is also not evident by drawing analogies from the description in the vibronic basis, and the same holds for the appearance of a ratio of Matsubara decomposition coefficients when the non-Condon transition dipole moment acts from the right hand side.
Note that our description of dependencies of transition dipole moments on vibrational coordinates by establishing connections to adjacent ADOs relies on a similar concept as the treatment of the respective effects with DEOM in Ref.~\cite{ZhQiXu16_JCP_204109}, where such dependencies are expressed in terms of so-called dissipatons. In \cite{ZhQiXu16_JCP_204109} also the quasi-particle nature of dissipatons was mentioned, which anticipates our independently developed interpretation of the ADOs as being connected to each other via creation and annihilation operators.

For the treatment of second-order Herzberg-Teller coupling with HEOM, we choose a heuristic approach instead of a rigorous derivation with the corresponding Hamiltonian from Eq.~(\ref{eq:nonadiabatic_coupling_second_order}) entering in the time evolution from Eq.~(\ref{eq:time_evolution_density_matrix}) and, via cumulant expansion, in the Feynman-Vernon functional. 
Such derivation, which would lead to appearance of higher-order correlation functions in the Feynman-Vernon functional due to involvement of second-order Herzberg-Teller contributions, would be rather cumbersome. Furthermore, including a bilinear term in the Feynman-Vernon functional is problematic because it leads to distortion of the thermal equilibrium state and of the normalization
of the density matrix elements. Approaches to overcome this issue by treating the single oscillator mode explicitly and either carrying out path integrations or solving the equation of motion for the total system have been discussed in the literature \cite{TaOk97_JCP_2018,TaTa09_JPSJ_073802}.
Our heuristic approach relies on identification of the way how a term with linear dependence on a vibrational coordinate, such as the first-order Herzberg-Teller coupling term, translates into involvement of adjacent ADOs in the respective contributions to HEOM. The extracted scheme of referring to adjacent ADOs can be formulated in terms of an operator acting in HEOM space, which is applied recursively two times for the treatment of second-order Herzberg-Teller coupling under the assumption that second-order Herzberg-Teller coupling contributions can be made accessible by generalizing the procedure for treatment of first-order Herzberg-Teller coupling.
Different from both the numerically exact standard formulation of HEOM and our additional first-order Herzberg-Teller coupling contributions, which are obtained by analytic derivation, our heuristic treatment of second-order Herzberg-Teller coupling might be of limited applicability. 
Although we checked the plausibility of our treatment of second-order Herzberg-Teller coupling by drawing connections to terms appearing in the framework of a rigorous derivation, a proof of the equivalence of both approaches seems not to be straightforward and would go beyond the scope of this article. Numerical results presented later in this paper confirm the validity of our treatment in the studied parameter regime.

In the recursion steps specified in the following, the second-order Herzberg-Teller coupling is rescaled by factors $\frac{\sqrt{c_k}}{\sqrt{S_{UO}} \omega_{UO}}$ and $\frac{\sqrt{c_{k'}}}{\sqrt{S_{UO}} \omega_{UO}}$ with assignment of the Matsubara decomposition indices $k$ and $k'$ to the involved 
vibrational coordinates, resulting in second-order couplings $\bar{J}_{12,kk'}^{(2)}$.
The recursion scheme can be obtained by considering the general form of a first-order Herzberg-Teller contribution to the HEOM scheme, as defined in Eq.~(\ref{eq:evaluation_HEOM_scheme_dimer_second_step_J1}), and by extracting the influence of the coupling operator in an analogous way as in the case of a non-Condon transition dipole moment. 
To account for Herzberg-Teller coupling of $M$-th order by applying the operator $\hat{V}^{(m)}_{H,J^{(M)}}$ with initial recursion index $m=M$ and termination condition $\hat{V}^{(0)}_{H,J^{(M)}}=\hat{V}_{J^{(M)}}$, we define the recursion scheme component-wise as
\begin{widetext}
\begin{eqnarray} \label{eq:influence_nonadiabatic_coupling_term_second_order_product1}
\hat{V}^{(m)}_{H,J^{(M)},i,j} \hat{\rho}_{\boldsymbol{n}'} 
= \sum_{k'} -\sqrt{\frac{1}{c_{k'}}} && \left(
\hat{V}^{(m-1)}_{H,J^{(M)},i,j} \hat{\rho}_{\boldsymbol{n}'^{+}_{ik'}}
+n'_{ik'} c_{k'} \hat{V}^{(m-1)}_{H,J^{(M)},i,j} \hat{\rho}_{\boldsymbol{n}'^{-}_{ik'}} \right. \nonumber \\
&&\left. +\hat{V}^{(m-1)}_{H,J^{(M)},i,j} \hat{\rho}_{\boldsymbol{n}'^{+}_{jk'}}
n'_{jk'} c_{k'} \hat{V}^{(m-1)}_{H,J^{(M)},i,j} \hat{\rho}_{\boldsymbol{n}'^{-}_{jk'}} \right), \\
\hat{\rho}_{\boldsymbol{n}'} \hat{V}^{(m)}_{H,J^{(M)},i,j} 
= \sum_{k'} -\sqrt{\frac{1}{c_{k'}}} && \left( 
\hat{\rho}_{\boldsymbol{n}'^{+}_{ik'}} \hat{V}^{(m-1)}_{H,J^{(M)},i,j}
+n'_{ik'} \tilde{c}_{k'} \hat{\rho}_{\boldsymbol{n}'^{-}_{ik'}} \hat{V}^{(m-1)}_{H,J^{(M)},i,j}  \right. \nonumber \\
&&+\hat{\rho}_{\boldsymbol{n}'^{+}_{jk'}} \hat{V}^{(m-1)}_{H,J^{(M)},i,j}
\left. +n'_{jk'} \tilde{c}_{k'} \hat{\rho}_{\boldsymbol{n}'^{-}_{jk'}} \hat{V}^{(m-1)}_{H,J^{(M)},i,j} \right).
\end{eqnarray}
\end{widetext}
With these definitions, the contribution of second-order Herzberg-Teller coupling in the framework of HEOM can be written as
\begin{equation} \label{eq:evaluation_HEOM_scheme_J2_first_order}
\left( \frac{\partial}{\partial t} \hat{\rho}_{\boldsymbol{n}} \right)_{\bar{J}^{(2)}}
= -i \sum_i \sum_{j \neq i}  \frac{\tilde{J}^{(2)}_{ij}}{2} \left[ \hat{V}^{(2)}_{H,J^{(2)},i,j},\hat{\rho}_{\boldsymbol{n}} \right].
\end{equation}
For a compact formulation of the resulting expression we extend the notation from Eq.~(\ref{eq:redefinition_commutators}) by
\begin{equation} \label{eq:redefinition_commutators2}
\left[ \bullet,\hat{\tilde{\rho}}_{\boldsymbol{n}^{- \; \; \; -}_{ik \; jk'}} \right]_c=\bullet \hat{\tilde{\rho}}_{\boldsymbol{n}^{- \; \; \; -}_{ik \; jk}} 
-\frac{\tilde{c}_k}{c_k} \frac{\tilde{c}_{k'}}{c_{k'}} \hat{\tilde{\rho}}_{\boldsymbol{n}^{- \; \; \; -}_{ik \; jk'}} \bullet,
\end{equation}
where the scaling of the subtracted term with a product of ratios of Matsubara decomposition coefficients only appears in the case of two decreases of index digits in the involved ADO. If only one of the index digits decreases, the definition from Eq.~(\ref{eq:redefinition_commutators}) is used. After rescaling of the ADOs, the resulting expression is
\begin{widetext}
\begin{equation} \label{eq:coupling_second_order_HEOM_analytic_derivation}
\begin{split}
&\left( \frac{d}{dt} \hat{\tilde{\rho}}_{\boldsymbol{n}} \right)_{J^{(2)}}
=-i \sum_i \sum_{j \neq i} \sum_k \sum_{k'} \frac{1}{2} \bar{J}_{ij,kk'}^{(2)}
\Bigg( \Bigg[ \hat{B}^{\dagger}_i \hat{B}_j,
\Bigg( \\
&\delta_{kk'} \Bigg( \sqrt{n_{ik}+2} \sqrt{n_{ik'}+1} \hat{\tilde{\rho}}_{\boldsymbol{n}^{+ \; \; \; +}_{ik \; ik'}}
+\sqrt{n_{ik}+1} \sqrt{n_{ik'}+1} \hat{\tilde{\rho}}_{\boldsymbol{n}}
+\sqrt{n_{ik}} \sqrt{n_{ik'}} \hat{\tilde{\rho}}_{\boldsymbol{n}}
+\sqrt{n_{ik}-1} \sqrt{n_{ik'}} \hat{\tilde{\rho}}_{\boldsymbol{n}^{- \; \; \; -}_{ik \; ik'}} \\
&+ \sqrt{n_{jk}+2} \sqrt{n_{jk'}+1} \hat{\tilde{\rho}}_{\boldsymbol{n}^{+ \; \; \; +}_{jk \; jk'}}
+ \sqrt{n_{jk}+1} \sqrt{n_{jk'}+1} \hat{\tilde{\rho}}_{\boldsymbol{n}}
+ \sqrt{n_{jk}} \sqrt{n_{jk'}} \hat{\tilde{\rho}}_{\boldsymbol{n}}
+ \sqrt{n_{jk}-1} \sqrt{n_{jk'}} \hat{\tilde{\rho}}_{\boldsymbol{n}^{- \; \; \; -}_{jk \; jk'}} \Bigg) \\
&+(1-\delta_{kk'}) \Bigg( \sqrt{n_{ik}+1} \sqrt{n_{ik'}+1} \hat{\tilde{\rho}}_{\boldsymbol{n}^{+ \; \; \; +}_{ik \; ik'}}
+\sqrt{n_{ik}+1} \sqrt{n_{ik'}} \hat{\tilde{\rho}}_{\boldsymbol{n}^{+ \; \; \; -}_{ik \; ik'}} 
+\sqrt{n_{ik}} \sqrt{n_{ik'+1}} \hat{\tilde{\rho}}_{\boldsymbol{n}^{- \; \; \; +}_{ik \; ik'}} \\
&+\sqrt{n_{ik}-1} \sqrt{n_{ik'-1}} \hat{\tilde{\rho}}_{\boldsymbol{n}^{- \; \; \; -}_{ik \; ik'}}
+ \sqrt{n_{jk}+1} \sqrt{n_{jk'}+1} \hat{\tilde{\rho}}_{\boldsymbol{n}^{+ \; \; \; +}_{jk \; jk'}}
+ \sqrt{n_{jk}+1} \sqrt{n_{jk'}} \hat{\tilde{\rho}}_{\boldsymbol{n}^{+ \; \; \; -}_{jk \; jk'}} \\
&+ \sqrt{n_{jk}} \sqrt{n_{jk'}+1} \hat{\tilde{\rho}}_{\boldsymbol{n}^{- \; \; \; +}_{jk \; jk'}} 
+ \sqrt{n_{jk}-1} \sqrt{n_{jk'}-1} \hat{\tilde{\rho}}_{\boldsymbol{n}^{- \; \; \; -}_{jk \; jk'}} \Bigg) \\
&+ 2 \Bigg( \sqrt{n_{ik}+1} \sqrt{n_{jk'}+1} \hat{\tilde{\rho}}_{\boldsymbol{n}^{+ \; \; \; +}_{ik \; jk'}}
+ \sqrt{n_{ik}+1} \sqrt{n_{jk'}} \hat{\tilde{\rho}}_{\boldsymbol{n}^{+ \; \; \; -}_{ik \; jk'}} 
+ \sqrt{n_{ik}} \sqrt{n_{jk'}+1} \hat{\tilde{\rho}}_{\boldsymbol{n}^{- \; \; \; +}_{ik \; jk'}} \\
&+ \sqrt{n_{ik}} \sqrt{n_{jk'}} \hat{\tilde{\rho}}_{\boldsymbol{n}^{- \; \; \; -}_{ik \; jk'}} \Bigg)
\Bigg) \Bigg]_c \Bigg).
\end{split}
\end{equation}
\end{widetext}
Note that because of taking two steps in the derivation (corresponding to selection of second-order terms in the system-bath interaction),
where each step implies an inverted sign (as compared to the usual sign of excitonic coupling contributions in HEOM equations), the sign changes compensate each other.
Apart from this aspect, the previous considerations about the question to which extent drawing analogies from the vibronic basis description of the first-order Herzberg-Teller coupling term to its description in the HEOM formalism is appropriate, also apply in the case of the second-order Herzberg-Teller coupling term with formulation in the vibronic basis given in Eq.~(\ref{eq:J_second_order}).
First, the second-order Herzberg-Teller coupling elements are scaled by temperature-dependent factors.
Second, the subtracted terms in the commutators involving ADOs with decreasing values of index digits are scaled by ratios of Matsubara decomposition coefficients.
In addition to these aspects, in the second-order Herzberg-Teller coupling term one finds additional contributions with simultaneous changes of index digits which are assigned to different Matsubara decomposition terms, even if those index digits are related to the correlation function of the same monomer.
Such terms would not be expected by drawing analogies from the vibronic basis description, and they seem to be characteristic for the treatment of undamped oscillator contributions, where the combination of both Matsubara decomposition terms is required for a comprehensive description. 
All other terms can be easily assigned to corresponding terms in the vibronic basis representation and drawn back to the involvement of a product of specific position coordinates. In this way, it is possible to select, for example, contributions from squared position coordinates or of mixed products of them.
In the case of a Brownian spectral density for the description of an underdamped oscillator, where the two lowest Matsubara decomposition terms are similar to those of the undamped oscillator, simultaneous changes of index digits assigned to other combinations of Matsubara decomposition terms seem not to play a role.

Note that higher-order Herzberg-Teller coupling terms can be made accessible by applying an analogous recursion scheme, as exemplified for the second-order Herzberg-Teller coupling. Again, it is worth mentioning that such contributions are obtained from a heuristic approach and that a numerical exactness of the RDM evolution is not guaranteed if such terms become involved.

\subsection{Calculation of absorption spectra} \label{sec:absorption_spectra}

Both in the case of HEOM and of density matrix propagation in a vibronic basis, absorption spectra can be calculated from a correlation function including the transition dipole operators 
$\hat{\vec{\mu}}_{+}=\sum_m (\vec{\mu}^{(0)}_m+\vec{\mu}^{(1)}_m \hat{q}_m) \hat{B}_m^{\dagger}$ and $\hat{\vec{\mu}}_{-}=\sum_m (\vec{\mu}^{(0)}_m+\vec{\mu}^{(1)}_m \hat{q}_m) \hat{B}_m$, 
a Green operator ${\cal G}(t)$ to account for time propagation and the initial system density matrix $\hat{\rho}_0$.
The general expression for the correlation function reads 
\begin{equation} \label{eq:correlation_function_absorption}
C_{abs}(t)=\langle \hat{\vec{\mu}}_{-} {\cal G}(t) \hat{\vec{\mu}}_{+} \hat{\rho}_0 \rangle.
\end{equation}
Note that in the HEOM description $\langle \bullet \rangle$ denotes the trace over the system density matrix $\tilde{\rho}_{\boldsymbol{0}}(t)$ after application of $\hat{\vec{\mu}}_{-} {\cal G}(t) \hat{\vec{\mu}}_{+}$, while contributions of other ADOs are not taken into account, at least under Condon approximation.
If non-Condon effects are involved, also ADOs from the hierarchy layer with change of a single index digit by $+1$ compared to the system density matrix yield a contribution. 
Further ADOs from the hierarchy structure enter via time evolution, independent of whether non-Condon effects are taken into account or not.
The absorption spectrum is obtained from the correlation function by Fourier transformation according to the formula

\begin{equation}
\sigma_{abs}(\omega)=\int^{\infty}_{0} \exp(-i \omega t) C_{abs}(t).
\end{equation}

To account for orientational averaging effects in the calculation of absorption spectra, the respective correlation function can be reformulated as
\begin{equation} \label{eq:correlation_function_absorption_separation_pathways}
\begin{split}
C_{abs}(t)&=\langle \hat{\vec{\mu}}_{-} \hat{\cal G}(t) \hat{\vec{\mu}}_{+} \hat{\rho}_0 \rangle \nonumber \\
&=\sum_i \sum_j \langle \hat{\vec{\mu}}_{-,i} \hat{\cal G}_{ij}(t) \hat{\vec{\mu}}_{+,j} \hat{\rho}_0 \rangle.
\end{split}
\end{equation}
For description in the vibronic eigenstate basis with treatment of dissipation effects by the secular Redfield approach only contributions with $i=j$ are obtained, whereas in the case of HEOM also non-secular terms with $i \neq j$ can appear.
If such terms are neglibile, orientational averaging according to \cite{HeKrKr12_NJP_023018} only leads to a scaling factor without changing the appearance of the absorption spectrum.

\section{Results} \label{sec:results}

\begin{figure}[H] 
\begin{center}
\includegraphics*[width=0.75\columnwidth]{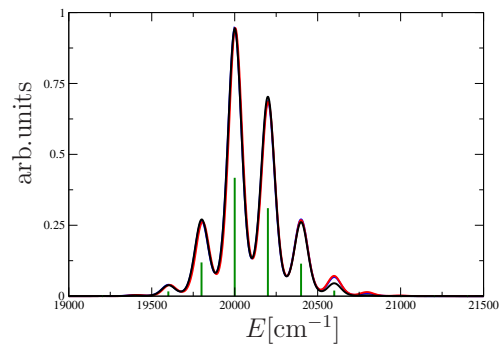}
\end{center}
\caption{Monomer absorption spectra calculated with density matrix propagation in vibronic basis (black line: five vibrational levels, blue line: seven vibrational levels) and HEOM (red line) are shown together with stick spectra (green line).
Parameters: $\epsilon_M=\unit[20000]{cm^{-1}}$, $\omega_{0}=\unit[200]{cm^{-1}}$, $S_M=\unit[0.5]{}$.}
\label{fig:absorption_spectrum_density_matrix_propagation_without_excitonic_coupling}
\end{figure}

\begin{figure}[H] 
\begin{center}
\includegraphics*[width=0.75\columnwidth]
{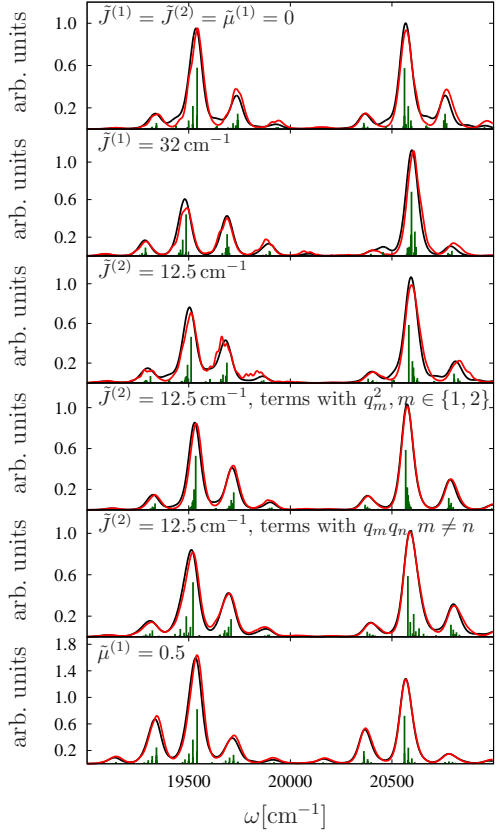}
\end{center}
\caption{Dimer absorption spectra, calculated with density matrix propagation in vibronic basis (black lines) and HEOM (red lines) together with stick spectra (green line). In the first row the results without involvement of Herzberg-Teller coupling- and non-Condon effects are shown. From the second row downwards a single Herzberg-Teller coupling- or non-Condon parameter is taken as non-zero, as specified by the label of each subfigure.}
\label{fig:absorption_spectra}
\end{figure}

\begin{figure}[H] 
\begin{center}
\includegraphics*[width=0.75\columnwidth]{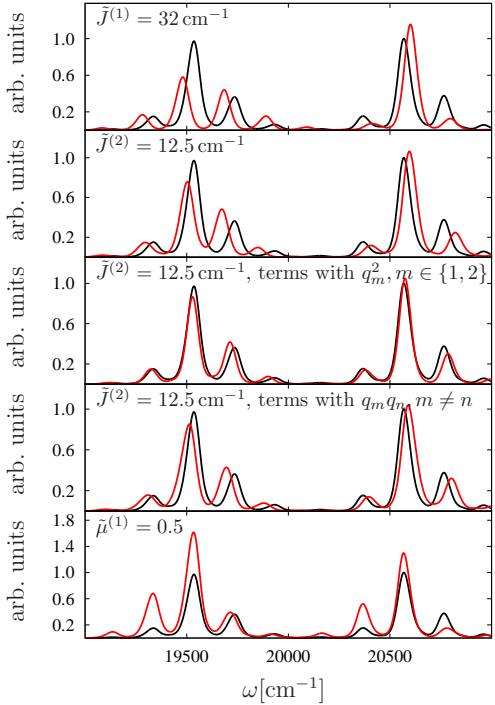}
\end{center}
\caption{Dimer absorption spectra, calculated with density matrix propagation in vibronic basis: Comparison of results for $\tilde{J}_{12}^{(1)}=\unit[0]{cm^{-1}}$, $\tilde{J}_{12}^{(2)}=\unit[0]{cm^{-1}}$ and $\tilde{\mu}^{(1)}=\unit[0]{}$ (black lines) with results from calculations with a single Herzberg-Teller coupling- and non-Condon parameter taken as non-zero, as specified by the label of each subfigure (red lines).}
\label{fig:absorption_spectra_comparison_different_cases}
\end{figure}

For a comparison of absorption spectra from density matrix propagation in the vibronic basis and those from HEOM propagation,
we choose a dimer model system with the following parameters: 
The monomer units are taken as equal, with electronic excitation energies $\epsilon_1=\epsilon_2=\unit[20000]{cm^{-1}}$
and excitonic coupling $J_{12}=\unit[500]{cm^{-1}}$. The intramolecular vibrational modes are characterized by the vibrational frequencies 
$\omega_{1}=\omega_{2}=\unit[200]{cm^{-1}}$ and the Huang-Rhys factors $S_1=S_2=\unit[0.5]{}$.
The angle between the transition dipole moments of equal absolute value is taken as $\theta=\unit[90]{^\circ}$. Therefore, the matrix elements of the transition dipole operator $\hat{\vec{\mu}}^{(0)}$ in the exciton basis, which scale with $2 \cos^2(\theta/2)$ and $2 \sin^2(\theta/2)$, depending on the exciton state involved in the respective transition, lead to equal transition dipole amplitudes for both exciton bands. The same scaling applies for the non-Condon contribution $\hat{\vec{\mu}}^{(1)}$.
In the Debye-Drude spectral density component we use the parameter values $\eta_{DD}=0.56$ and $\omega_c=\unit[20]{cm^{-1}}$.
To study the influence of first- and second-order Herzberg-Teller coupling and of the non-Condon transition dipole contributions separately, only one of the corresponding parameters is always taken as non-zero, and the results are compared to the case where all of them are equal to zero.
Note that because of $q_0=\sqrt{\frac{2}{\omega_0}}$ (see Ref.~\cite{DoMaVa14_JCP_115103}) the parameters $\tilde{\mu}^{(1)}$, $\tilde{J}_{12}^{(1)}$ and $\tilde{J}_{12}^{(2)}$, multiplied by a factor of $\unit[2]{}$ (first-order terms) and $\unit[4]{}$ (second-order terms), respectively, can be considered as proportionality constants of Herzberg-Teller coupling or non-Condon transition dipole contribution with respect to the rescaled, dimensionless vibrational coordinate $\frac{q_i}{q_0}$ (first-order terms) or products of such dimensionless coordinates (second-order term). 
We assume $\tilde{J}_{12}^{(1)}=\unit[32]{cm^{-1}}$, $\tilde{J}_{12}^{(2)}=\unit[12.5]{cm^{-1}}$ and $\tilde{\mu}^{(1)}=\unit[0.5]{}$, but select only one of these parameters to be different from zero in each calculation to be able to study its influence separately. 
Together with the absorption spectra we show stick spectra, which are obtained by assigning a stick with the length corresponding to the square of the transition dipole moment between the energy eigenstates obtained by diagonalization of the system Hamiltonian in the vibronic basis.
First, we consider monomer spectra from density matrix propagation in a vibronic basis and from HEOM, which are expected to be identical, provided that the numerical calculation is sufficiently accurate.
In this respect, among other parameters, the number of included vibrational eigenstates in the vibronic basis plays a role.
For five vibrational levels, one still finds small differences in the vibrational side bands at the largest energetic positions, as shown in Fig.~\ref{fig:absorption_spectrum_density_matrix_propagation_without_excitonic_coupling}.
%
For seven vibrational levels almost no differences in the absorption spectra from both calculation methods appear anymore.
However, with increasing number of vibrational levels also the numerical effort substantially increases, particularly in the treatment of the dimer, so that the default number of five vibrational levels can be considered as an appropriate choice for the given model parameters.
Also in the case with excitonic coupling under the assumption $\tilde{J}_{12}^{(1)}=\tilde{J}_{12}^{(2)}=\tilde{\mu}^{(1)}=\unit[0]{}$ (see Fig.~\ref{fig:absorption_spectra}, first row) one finds an extensive agreement of the results of the calculations by both methods.
The reduced relative intensities of the vibrational side bands can be explained by an effective Huang-Rhys factor of exciton state $\alpha$, which in the case of a homodimer with $S_1=\zeta_1 S_D=\frac{1}{2} S_D$, $S_2=\zeta_2 S_D=\frac{1}{2} S_D$, $S_D=S_1+S_2$ and $\langle \alpha | 1 \rangle=\langle \alpha | 2 \rangle=\sqrt{\frac{1}{2}}$ becomes $S_{\alpha}=\frac{1}{4} S_D=\unit[0.25]{}$, as described in the Supplementary Material. In Fig.~\ref{fig:absorption_spectra_effective_HR-factor} appropriately shifted and scaled absorption spectra of monomers with this Huang-Rhys factor exhibit very similar vibrational structure as the exciton bands of the dimer spectra.
\begin{figure}[H]
\begin{center}
\includegraphics*[width=0.75\columnwidth]{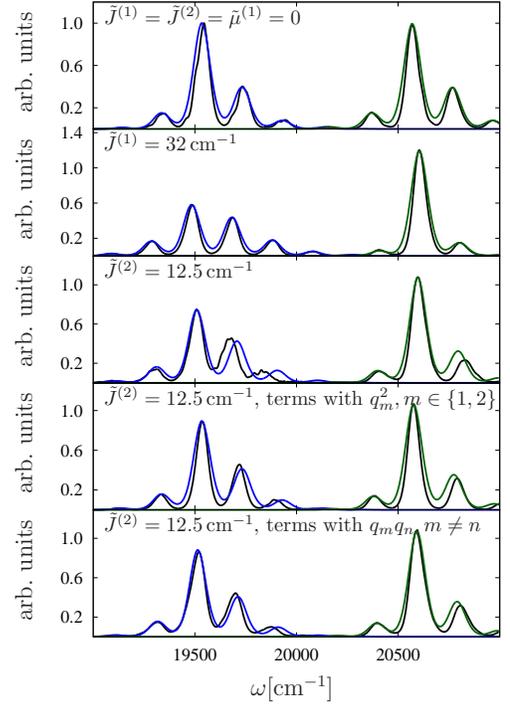}
\end{center}
\caption{Dimer absorption spectra from HEOM calculations (black lines) are shown together with monomer absorption spectra from HEOM calculations with effective Huang-Rhys factors $S_{\alpha_1}$ and $S_{\alpha_2}$ which allow to reproduce the vibrational peak progressions of upper and lower exciton band (green and blue lines, respectively) after adjusting position and height. In the first row the results without involvement of Herzberg-Teller coupling- and non-Condon effects are shown. From the second row downwards a single Herzberg-Teller coupling- or non-Condon parameter is taken as non-zero, as specified by the label of each subfigure.}
\label{fig:absorption_spectra_effective_HR-factor}
\end{figure}
For a non-zero first-order Herzberg-Teller coupling constant $\tilde{J}_{12}^{(1)}=\unit[32]{cm^{-1}}$ 
(see Fig.~\ref{fig:absorption_spectra}, second row)
the relative intensity of vibrational side bands is enhanced in the exciton band at lower energy, whereas in the upper exciton band it becomes less pronounced. 
This finding appears in the results of both calculation methods, which agree well in the positions and relative intensities of the peaks. 
A comparison of results from calculations in the vibronic basis with and without first-order Herzberg-Teller coupling effects, as displayed in the first row of Fig.~\ref{fig:absorption_spectra_comparison_different_cases}, shows that besides the change of the effective Huang-Rhys factor, also the splitting between the exciton bands is modified under the influence of $\tilde{J}_{12}^{(1)}$.
The related effective Huang-Rhys factors, which determine the vibrational peak progression of the exciton bands, can be obtained as follows:
Transformation of both the system-bath coupling contributions and the first-order Herzberg-Teller coupling contribution of the Hamiltonian to the exciton basis leads to 
\begin{widetext}
\begin{equation}
\begin{split}
S_{\alpha,eff,J^{(1)}}&=\frac{1}{2} \Bigg( 
\sqrt{S_D} (\sqrt{\zeta_1} |\langle \alpha | 1 \rangle|^2 + \sqrt{\zeta_2} |\langle \alpha | 2 \rangle |^2)
+\frac{2 \sum_l \sum_{m \neq l} \tilde{J}_{lm}^{(1)} \langle \alpha | l \rangle \langle m | \alpha \rangle }{\omega_0} \Bigg)^2 \\
&+\frac{1}{2} S_D (\sqrt{\zeta_1} |\langle \alpha | 1 \rangle|^2 - \sqrt{\zeta_2} |\langle \alpha | 2 \rangle |^2)^2,
\end{split}
\end{equation}
\end{widetext}
according to the derivation in the Supplementary Material. 
In the case of a homodimer one obtains effective Huang-Rhys factors $S_{\alpha_1,J^{(1)}}=\unit[0.525]{}$ and $S_{\alpha_2,J^{(1)}}=\unit[0.076]{}$ for energetically higher and lower exciton state, respectively. 
Separate calculations of monomer absorption spectra under the assumption of these Huang-Rhys factors allow us to reproduce the vibrational peak progression of the exciton bands of the dimer-spectra with non-zero first-order Herzberg-Teller coupling, as shown in the first row of Fig.~\ref{fig:absorption_spectra_comparison_different_cases}.

\begin{figure}[H]
\begin{center}
\includegraphics*[width=0.75\columnwidth]{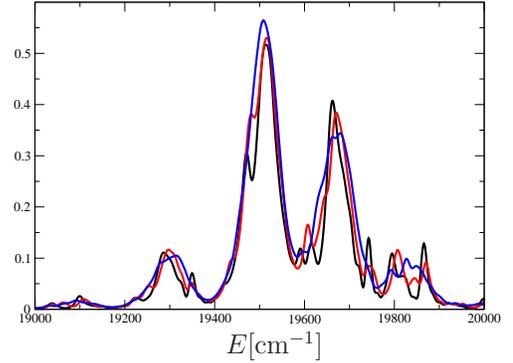}
\end{center}
\caption{For $\tilde{J}_{12}^{(1)}=\unit[0]{cm^{-1}}$, $\tilde{J}_{12}^{(2)}=\unit[12.5]{cm^{-1}}$ and $\tilde{\mu}^{(1)}=\unit[0]{}$ dimer absorption spectra from HEOM calculations with different truncation orders $N$ are compared in the region of the lower exciton band. The black, red and blue curve correspond to $N=\unit[10]{}$, $N=\unit[12]{}$ and $N=\unit[18]{}$, respectively.}
\label{fig:absorption_spectra_influence_truncation_order_and_damping_undamped_oscillator}
\end{figure}

\begin{figure}[H] 
\begin{center}
\includegraphics*[width=0.75\columnwidth]{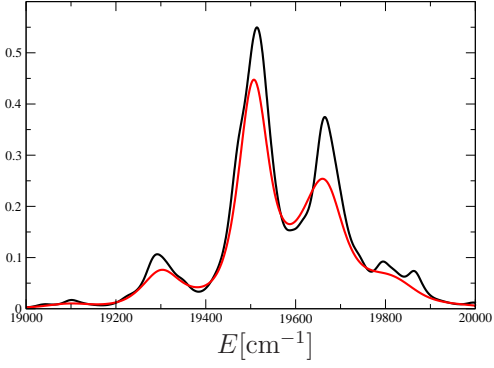}
\end{center}
\caption{Dimer absorption spectra for underdamped oscillator with damping constant $\gamma=\unit[5]{cm^{-1}}$ instead of undamped oscillator, but otherwise identical parameters as in the calculations of the spectra shown in Fig.~\ref{fig:absorption_spectra_influence_truncation_order_and_damping_undamped_oscillator}.
Results are only shown for $N=\unit[10]{}$ and $N=\unit[12]{}$ because sufficient numerical convergence is already achieved at the latter truncation order.}
\label{fig:absorption_spectra_influence_truncation_order_and_damping_underdamped_oscillator}
\end{figure}

For a non-zero second-order Herzberg-Teller coupling term $\tilde{J}_{12}^{(2)}=\unit[12.5]{cm^{-1}}$ a comparison of the results from density matrix propagation in the vibronic basis and from HEOM is displayed in the third row of Fig.~\ref{fig:absorption_spectra}.
Again, the Herzberg-Teller coupling enhances the vibrational peak progression of the lower exciton band and diminishes the one of the upper exciton band.
The vibrational peaks in the lower exciton band of the absorption spectrum calculated with HEOM are not as smooth as those in the absorption spectrum from density matrix propagation.
It seems that this finding can be explained by the numerical accuracy of the calculation, as increasing the truncation order in the HEOM calculation leads to better agreement between the
results from the different methods (see Fig.~\ref{fig:absorption_spectra_influence_truncation_order_and_damping_undamped_oscillator}). 
The assumption of an underdamped oscillator with description by a Brownian oscillator spectral density instead of the assumption of an undamped oscillator (see Fig.~\ref{fig:absorption_spectra_influence_truncation_order_and_damping_underdamped_oscillator}) leads to convergence already for a truncation order $N=\unit[12]{}$ of the Matsubara expansion, whereas in the case of an undamped oscillator the result for $N=\unit[12]{}$ is still remarkably different from the one for $N=\unit[18]{}$. 
We nevertheless keep using the undamped oscillator model because of its appropriateness for drawing analogies to the vibronic basis description on the level of the equations and because of the better comparability of the respective results.
Comparison of the absorption spectra with second-order Herzberg-Teller coupling and without any Herzberg-Teller coupling effects from calculations in the vibronic basis shows that the splitting of the exciton bands is not independent of their vibrational substructure (see Fig.~\ref{fig:absorption_spectra_comparison_different_cases}, second row).
This effect becomes recognizable more clearly if $J^{(2)}$-contributions of the Hamiltonian with involvement of either a product of position operators assigned to vibrational coordinates of different monomer units or of squared position operators assigned to the vibrational coordinate of the same monomer unit are considered separately (see Fig.~\ref{fig:absorption_spectra_comparison_different_cases}, third and fourth row). 
The comparison between the different calculation methods in the case where only contributions of the first type are taken into account leads to the absorption spectra shown in the fourth row of
Fig.~\ref{fig:absorption_spectra}, where a better agreement of the results than in the row above appears and numerical convergence is obtained for smaller truncation order. 
Similar agreement of the results appears in the case where only the contributions from terms with squared position coordinates are taken into account (see Fig.~\ref{fig:absorption_spectra}, fifth row), however numerical convergence requires a larger truncation order than in the previous case.
The differences in the absorption spectra from the compared calculation approaches in the case that both kinds of terms are included appear because of the increased numerical demands posed by their combination. For the numerical convergence the effective strength of the Herzberg-Teller coupling, which is obviously increased in a combination of both contributions, plays a role. However, also the interplay between both contributions seems to influence the numerical convergence. 
In the case of the second-order Herzberg-Teller coupling the influence on an effective displacement is more difficult to determine than in the case of the first-order Herzberg-Teller coupling. However, the situation can be simplified by assuming an averaged value of the vibrational coordinate of monomer $j$ which corresponds to the respective displacement $d_j$ when this monomer is excited. Furthermore, in the case of strong zeroth-order excitonic coupling one can assume that also the vibrational coordinate of the de-excited monomer $k \neq j$ is displaced by $d_k$ on average. The similar changes of the relative vibrational peak intensities in the contributions with involvement of only squared vibrational coordinates or only mixed products of them confirms this assumption. The effective Huang-Rhys factor under the influence of second-order Herzberg-Teller coupling, where scaling factors $f_k$ account for the influence of the Herzberg-Teller coupling on the displacement, can be identified as 
\begin{widetext}
\begin{equation}
\begin{split}
S_{\alpha,eff,J^{(2)}}&=\frac{1}{2} S_D \Bigg(
(\sqrt{\zeta_1} |\langle \alpha | 1 \rangle|^2 + \sqrt{\zeta_2} |\langle \alpha | 2 \rangle |^2)
+\frac{2 \sum_{k} f_k \sqrt{\zeta_k} \sum_l \sum_{m \neq l} \tilde{J}_{lm}^{(2)} \langle \alpha | l \rangle \langle m | \alpha \rangle }{\omega_0} \Bigg)^2 \\
&+\frac{1}{2} S_D (\sqrt{\zeta_1} |\langle \alpha | 1 \rangle|^2 - \sqrt{\zeta_2} |\langle \alpha | 2 \rangle |^2)^2,
\end{split}
\end{equation}
\end{widetext}
according to the derivation in the Supplementary Material. For the sake of simplicity we disregard an influence of Herzberg-Teller coupling on the displacement by setting $f_1=f_2=1$.
Accordingly, for the dimer model with the specified parameters effective Huang-Rhys factors of $S_{\alpha_1,J^{(2)}}=\unit[0.346]{}$ and $S_{\alpha_2,J^{(2)}}=\unit[0.169]{}$ for energetically higher and lower exciton state are obtained, respectively, if all second-order Herzberg-Teller coupling terms are taken into account. Again, separately calculated monomer absorption spectra for such Huang-Rhys factors allow to approximately reproduce the vibrational peak progression of the bands in the dimer-spectra with non-zero second-order Herzberg-Teller coupling, as shown in the second row of Fig.~\ref{fig:absorption_spectra_comparison_different_cases}.
The same holds for the separate terms shown in the third and fourth row of Fig.~\ref{fig:absorption_spectra_comparison_different_cases}, where the effective Huang-Rhys factors are $S_{\alpha_1,J^{(2)},mixed}=S_{\alpha_1,J^{(2)},squared}=\unit[0.296]{}$ and $S_{\alpha_2,J^{(2)},mixed}=S_{\alpha_2,J^{(2)},squared}=\unit[0.208]{}$.
If non-Condon effects are taken into account by assuming $\tilde{\mu}^{(1)}=\unit[0.5]{}$ 
(see Fig.~\ref{fig:absorption_spectra}, sixth row), 
the positions of the vibrational peaks of the exciton bands in absorption spectra calculated with density matrix propagation and HEOM agree well, and the relative intensities of the peaks differ only slightly. In both exciton bands the vibrational structure exhibits a redistribution of relative peak intensities, which leads to enhanced vibrational peaks at the bottom of the frequency range of each exciton band.
However, the peak positions of the vibrational sub-bands are not changed compared to the case without Herzberg-Teller coupling- and non-Condon effects, as the fifth row of Fig.~\ref{fig:absorption_spectra_comparison_different_cases} shows.
In the absorption spectra in the sixth row of Fig.~\ref{fig:absorption_spectra} a change of the relative intensity of the exciton bands under the influence of non-Condon transitions becomes recognizable. This effect seems to be caused by interference of the dependencies of the transition dipole moments on vibrational coordinates via the zeroth-order excitonic coupling, which also influence the excited state dynamics, as discussed in \cite{ZhQiXu16_JCP_204109}. However, such interference of monomer excitations beyond the Condon approximation via zeroth-order excitonic coupling cannot be identified with a dependence of the excitonic coupling itself on the position coordinates.

In all of the presented absorption spectra non-secular effects are negligible, so that orientational averaging simply results in an overall scaling factor and is therefore neglected. 
With increasing gap between the electronic excitation energies non-secular effects, which in the case of linear absorption correspond to coherence transfer contributions, gain relevance. Their influence becomes recognizable when the energy gap is adjusted to match the vibrational frequency -- a case of interest in the context of vibronic enhancement of excitation energy transport \cite{ChKaPuMa12_JPCB_7449,TiPeJo17_JCP_154308}.
Furthermore, when the excitonic coupling is reduced, so that the exciton bands get less separated, electronic excitation of one exciton state does not only lead to involvement of the vibrational substructure of the selected exciton band, but also of the vibrational substructure of the complementary band to a non-negligible extent. Both of the latter aspects bias the validity of the concept of an effective Huang-Rhys factor, which nevertheless remains useful for interpretation of the absorption spectra.

\section{Conclusions} \label{sec:conclusions}
In this work, we have developed a method for the description of Herzberg-Teller- and non-Condon effects in the framework of the HEOM for the reduced density matrix. It turned out that additional terms in the HEOM scheme, required to account for such effects, have similar structure as the related contributions to the Hamiltonian in the vibronic basis representation. However, the signs of the respective terms and the appearance of temperature-dependent factors in the HEOM description cannot be explained by drawing analogies from the vibronic basis description. Rather, an analytic derivation via a formulation in terms of path integrals with the Feynman-Vernon functional is required for a comprehensive treatment. A comparison of dimer absorption spectra calculated using both approaches shows an extensive agreement. The advantage of using HEOM instead of density matrix propagation in the vibronic basis for the treatment of Herzberg-Teller- and non-Condon effects lies in a more advantageous scaling of the numerical effort with increasing aggregate size, provided that an appropriate truncation scheme is used, and in the simultaneous numerically exact treatment of the thermodynamic bath. While Herzberg-Teller effects can lead to an increase of the effective Huang-Rhys factor in the singly excited state and impede the numerical convergence in this way, they leave the structure of the hierarchy unchanged. In the discussion of the absorption spectra we quantified the respective influence of Herzberg-Teller effects and applied the concept of an effective Huang-Rhys factor for the case of a homodimer. We found that Herzberg-Teller effects result in a redistribution of the oscillator strengths in the vibrational substructures of equally excited exciton bands, where in one exciton band the vibrational progression is enhanced, while in the other it is diminished. We also discussed the reliability of predictions using the effective Huang-Rhys factor under different model assumptions and pointed out under which conditions coherence transfer effects, which appear in HEOM description, but not in the vibronic basis treatment with secular approximation, become relevant. 
The investigation of excited state dynamics under the influence of Herzberg-Teller couplings, which is accessible by spectroscopic techniques beyond linear absorption, in particular by two-dimensional electronic spectroscopy, will be a part of our forthcoming work.

\section*{Acknowledgments}
This work was supported by the Czech Science Foundation (GACR) grant no. 17-22160S. For the HEOM calculations the ``Rostock HEOM package'' by Marco Schr\"{o}ter was extended to account for the involvement of Herzberg-Teller- and non-Condon effects. Access to computing and storage facilities owned by parties and projects contributing to the National Grid Infrastructure MetaCentrum, provided under the program ``Projects of Large Infrastructure for Research, Development, and Innovations'' (LM2010005) is highly appreciated.

\clearpage


%
%




\end{document}